\documentclass[12pt,preprint]{aastex}

\catcode`\@=11
\def\gapprox{\mathrel{\mathpalette\@versim>}}
\def\lapprox{\mathrel{\mathpalette\@versim<}}
\def\@versim#1#2{\lower2.45pt\vbox{\baselineskip0pt\lineskip0.9pt
     \ialign{$\m@th#1\hfil##\hfil$\crcr#2\crcr\sim\crcr}}}
\def\HI{{\ion{H}{1}}\ }
\def\HII{{\ion{H}{2}}\ }
\def\SII{{\ion{S}{2}}\ }
\def\NII{{\ion{N}{2}}\ }
\def\OI{{\ion{O}{1}}\ }
\def\OII{{\ion{O}{2}}\ }
\def\OIII{{\ion{O}{3}}\ }
\def\NeIII{{\ion{Ne}{3}}\ }

\catcode`\@=12

\newcommand{\OmM}{\ifmmode {\Omega_{\rm M}}\else $\Omega_{\rm M}$\fi}
\newcommand{\OmL}{\ifmmode {\Omega_{\Lambda}}\else $\Omega_{\Lambda}$\fi}
\newcommand{\kmps}{\ifmmode {\rm\,km~s^{-1}} \else ${\rm\,km\,s^{-1}}$\fi}

\received{}
\begin{document}

\title{Supernova Remnants and the Interstellar Medium of M83: \newline Imaging \& Photometry with the WFC3 on HST.}

\author{Michael~A.~Dopita \altaffilmark{1}, 
William~P.~Blair \altaffilmark{2}, 
Knox~S.~Long \altaffilmark{3}, 
Max~Mutchler \altaffilmark{3}, 
Bradley~C.~Whitmore \altaffilmark{3}, 
Kip~D.~Kuntz \altaffilmark{2}, 
Bruce~Balick \altaffilmark{4}, 
Howard~E.~Bond \altaffilmark{3}, 
Daniela~Calzetti \altaffilmark{5}, 
Marcella~Carollo \altaffilmark{6}, 
Michael~Disney \altaffilmark{7}, 
Jay~A.~Frogel \altaffilmark{8}, 
Robert O'Connell \altaffilmark{9}, 
Donald~Hall \altaffilmark{10}, 
Jon~ A.~Holtzman \altaffilmark{11}, 
Randy~A.~Kimble \altaffilmark{12}, 
John MacKenty \altaffilmark{3}, 
Patrick~McCarthy \altaffilmark{13}, 
Francesco~Paresce \altaffilmark{14}, 
Abhijit~Saha \altaffilmark{15}, 
Joe~Silk \altaffilmark{16}, 
Marco~Sirianni \altaffilmark{17}, 
John~Trauger \altaffilmark{18}, 
Alistair~R.~Walker \altaffilmark{15}, 
Rogier~Windhorst \altaffilmark{19}, 
Erick~Young  \altaffilmark{20}}

\email{Michael.Dopita@anu.edu.au}
\altaffiltext{1}{Research School of Astronomy \& Astrophysics, The Australian National University, Cotter Road, \newline Weston Creek, ACT 2611, Australia}
\altaffiltext{2}{Johns Hopkins University, Baltimore, MD, USA}
\altaffiltext{3}{Space Telescope Science Institute, Baltimore, MD, USA}
\altaffiltext{4}{Dept. of Astronomy, University of Washington, Seattle, WA 98195-1580, USA}
\altaffiltext{5}{Dept. of Astronomy, University of Massachusetts, Amherst, MA 01003, USA}
\altaffiltext{6}{Institute of Astronomy, ETH-Zurich, Zurich, 8093 Switzerland}
\altaffiltext{7}{Dept. of Physics and Astronomy, Cardiff University,Cardiff CF24 3AA, United Kingdom}
\altaffiltext{8}{Association of Universities for Reserach in Astronomy., Washington, DC 20005, USA}
\altaffiltext{9}{Dept. of Astronomy, University of Virginia, Charlottesville, VA 22904-4325, USA}
\altaffiltext{10}{Institute for Astronomy, Honolulu, HI 96822, USA}
\altaffiltext{11}{New Mexico State University, Las Cruces, NM 88003, USA}
\altaffiltext{12}{Goddard Space Flight Center, Greenbelt, MD 20771, USA}
\altaffiltext{13}{Carnegie Institute of Washington, Pasadena, CA 91101-1292, USA}
\altaffiltext{14}{Institute of Space Astrophysics, INAF, 40129 Bologna, Italy}
\altaffiltext{15}{NOAO, Tucson, AZ 85726-6732, USA}
\altaffiltext{16}{Dept. of Physics, University of Oxford, Oxford OX1 3PU, United Kingdom}
\altaffiltext{17}{European Space Agency, Darmstadt, D-64293, Germany}
\altaffiltext{18}{NASA JPL, Pasadena, CA 91109, USA}
\altaffiltext{19}{Arizona State University, Tempe, AZ 85287-1404, USA}
\altaffiltext{20}{University of Arizona, Tucson, AZ 85721-0065, USA}


\begin{abstract}
We present Wide Field Camera 3 images taken with the Hubble Space Telescope within a single field in the southern grand design star-forming galaxy M83.  Based on their size, morphology and photometry in continuum-subtracted H$\alpha$, [\SII], H$\beta$, [\OIII] and [\OII] filters, we have identified 60 supernova remnant candidates, as well as a handful of young ejecta-dominated candidates. A catalog of these remnants, their sizes and, where possible their H$\alpha$ fluxes are given. Radiative ages and pre-shock densities are derived from those SNR which have good photometry. The ages lie in the range $2.62 < log(\tau_{\rm rad}/{\rm yr}) < 5.0$, and the pre-shock densities at the blast wave range over $0.56 < n_0/{\rm cm^{-3}} < 1680$. Two populations of SNR have been discovered. These divide into a nuclear and spiral arm group and an inter-arm population. We infer an arm to inter-arm density contrast of 4. The surface flux in diffuse X-rays is correlated with the inferred pre-shock density, indicating that the warm interstellar medium is pressurised by the hot X-ray plasma. We also find that the interstellar medium in the nuclear region of M83 is characterized by a very high porosity and pressure and infer a SNR rate of one per 70-150~yr for the nuclear ($R<300~$pc) region. On the basis of the number of SNR detected and their radiative ages, we infer that the lower mass of Type II SNe in M83 is $M_{\rm min} = 16^{+7}_ {-5}$~M$_{\odot}$. Finally we give evidence for the likely detection of the remnant of the historical supernova, SN1968L.
\end{abstract}


\section{Introduction}\label{sec:intro}
The techniques for the detection of supernova remnants (SNR) in external galaxies were pioneered by \cite{Math64,Wes66} and \cite{Math73}. These techniques used the characteristic non-thermal radio spectrum produced by SNR along with their optical characteristics, especially the relative strength of their [\SII]$\lambda\lambda 6717,31$ lines relative to H$\alpha$. Later, \citet{Dop77a} established that these optical characteristics were a natural consequence of the radiative shock waves associated with the propagation of the supernova blast wave through the interstellar medium (ISM) of the host galaxy, and that the details of the radiative spectrum in turn depends on the chemical abundances in the ISM of the host galaxy, rather than on the chemical abundances in the stellar ejecta \citep{Dop77b}. Strong [\SII]$\lambda\lambda 6717,31$ and [\OII]$\lambda\lambda 3727,9 $ emission arises in the recombination zones of these radiative shocks, where much of the Balmer line emission is also produced.

The characteristic strength of the  [\SII]/H$\alpha$ ratio seen in  SNR (typically greater than 0.4) has proven to be an extraordinarily versatile technique for detecting regions excited by radiative shocks. The association of such regions with either non-thermal radio emission or X-rays emitted from the shock-heated gas behind the SNR blast wave has usually sufficed to confirm the identification of a SNR. This technique has been successfully used to survey for SNR in the Magellanic Clouds \citep{Lasker77, Math83,Math84,Math85}, in M31 and other local group galaxies \citep{DOd80}, and in M33 \citep{DOd78,Blair81,Long90,Gord98,Long09}.

M83 (NGC 5236) is a southern Grand-Design SAB(s)c galaxy. It has an inclination angle of 24 degrees and has an angular size of $\sim13$ \arcmin.  Its distance is  given in the NASA/IPAC Extragalactic Database (NED) as $4.56\pm 0.09$ Mpc using a combination of redshift-independent techniques. It has well-formed spiral arms and distinct dust lanes. Detailed observations at radio \citep{Crossthwaite02,Maddox06}, infrared\citep{Vogler05,Rubin07}, ultraviolet\citep{Bohlin90,Boisser05,Thilker05}, X-ray \citep{SoWu02,SoWu03,Kilgard05} and optical wavelengths \citep{Calzetti04} indicate that the nucleus and spiral arms of M83 are regions of intense star formation. \cite{DOd85} found absorption features in the nucleus consistent with a population dominated by massive OB stars, which are believed to be the precursors of Type Ib,c/ II SNe \citep{Huang87}. Observations by \citet{Jensen81} revealed a large number of young, massive stars along the inner arms, indicating active star formation in these regions as well. These observations are consistent with a recent and ongoing burst of star formation lasting over at least the last $\sim 10^7$ \citep{Trincheri85}. Indeed, it has been the site of no less than six optically-detected supernova events in the past century making it the most active galaxy in this respect along with NGC~6946. This ubiquitous star formation and its high surface density makes M83 a natural target for SNR studies. Indeed it has been the subject of a detailed study by  \citet{BL04}, who identified a total of 71 candidates over the whole face of M83 using the ratio of the [\SII]$\lambda\lambda 6717,31$\AA\ lines to H$\alpha$, and followed up with spectroscopy of 25 of these. They found that 15 of their candidates were also associated with soft X-ray sources in the catalog of \citet{SoWu03} or with non-thermal radio sources \citep{Maddox06}.

In this paper, we present new observations of a single field in M83 taken by Wide Field Camera 3 (WFC3) on the Hubble Space Telescope (HST). These images were obtained in many narrow band and broadband filters, and we use these data to discover new SNR candidates not only using the traditional [\SII]/H$\alpha$ ratio, but also by the relative strength of their  [\OII]$\lambda\lambda 3727,9$\AA\ emission. Twelve of the previously discovered \citet{BL04} SNR candidates lie within this field. The complete list of candidates is presented in \ref{sec:Candidates}, and in section \ref{sect:Interpret} we present the interpretation of these data on the basis of their measured sizes and H$\alpha$ fluxes.  At the distance of M83, the pixel size of WFC3 (0.0396 arc sec) corresponds to a spatial scale of $0.9$~pc, and the spatial resolution at H$\alpha$ (0.06 arc sec) to a distance of 1.7~pc. This ensures that even the smallest of our SNR candidates is adequately resolved spatially.

\section{WFC3 Observations}\label{sect:Obs}

\subsection{The WFC3 dataset}
The WFC3 data were obtained as part of WFC Science Oversight Committee (SOC) Early Release Science (ERS) program (program ID11360: Principal Investigator: Robert O'Connell, The University of Virginia). The M83 field, centered at RA=13:37:04.42 Dec=-29:51:28.0 (J2000) was chosen to cover the nuclear region as well as the inner spiral arm which extends out from the northern side of the nucleus and curves away towards the north east. A second, outer field will be obtained later as part of this program. Either three or four dithered exposures were taken in each narrow-band filter to remove cosmetic defects and assist in the cosmic ray removal. For this paper a sub-set of the data taken in the UVIS channel has been used. The observing log for these observations is given in Table \ref{tab:Obs}. 

\subsection{Reduction of Observations}
Following acquisition, the data was passed through the MultiDrizzle software \citep{Fruchter09} to produce a cosmic-ray cleaned, combined image using the ``drizzle" software originally described by \citet{Fruchter02}.  This delivers a combined, cosmic ray cleaned and distortion corrected data product with a built-in world coordinate system. 

In these images, there is a small offset between images taken at different visits to the same target. These were removed using the \textsf{imalign} task in the IRAF package. Two weighted wide-band images were combined and scaled to provide a continuum reference image for each narrow-band image; F336W and F438W for the [\OII] data, F438W and F555W for the H$\beta$ and [\OIII] images and F555W and F814W for the H$\alpha$ and [\SII] images. The weighting of the broadband images was determined by the offset of their bandpass in logarithmic wavelength with respect to the emission line in question.  It should be noted that the F555W filter also transmits the [\OIII] emission line. However, in M83 the strength of this line is generally very feeble. This results from the super-solar chemical abundances in the inner region of this galaxy, and the resulting low electron temperatures in the \HII\ regions. As a consequence the photometric effect of this leakage is very small.

The mean redshift of M83 is only 513~km~s$^{-1}$. This is sufficiently small to ensure that the line emission is not significantly shifted from the peak of the filter bandpasses. The narrow band filter bandpasses are also sufficiently wide to ensure that the likely velocity dispersion of the SNR shell ($\sim 200-400$~km~s$^{-1}$) is also comfortably accommodated by the filter.

Each scaled continuum image was subtracted from the relevant line image to produce a pure emission line image. The subtraction is not perfect, because the color terms for the relatively unreddened hot blue stars in the spiral arms are quite different from those of the cool, heavily reddened, old stellar population in the nuclear region. The scaling of the continuum images was adjusted to provide the best compromise. Under-subtracted or over-subtracted residuals are still present, however.

\section{SNR Candidates}\label{sec:Candidates}
\subsection{Disk SNR}
In what follows, we will distinguish between the supernova remnant (SNR) candidates in the disk region of M83, and the nuclear region characterized by very high star formation rates and generally large reddening. To provide a physical definition of what constitutes the nuclear region, we  draw a circle with a radius of 300 pc centered at the highly-reddened nuclear cluster which we find at coordinates RA=13:37:00.871 Dec= -29:51:55.97 (J2000). 

The selection techniques to find SNR candidates are somewhat different between the disk and the nuclear region. Candidates in the general field were identified by blinking the aligned and match-scaled H$\alpha$, [\SII] and [\OII] images. The SNR candidates appear relatively bright  on both the [\SII] and [\OII] images compared with H$\alpha$. For these candidates, we could check the SNR identification by direct measurement of the  ratio of  [\SII]  to H$\alpha$ taking into account the fact that the HST F657N filter also transmits the [\NII] $\lambda 6584$ line. The ``classical'' definition of a SNR is [\SII]/H$\alpha > 0.4$. Since the [\NII] line may be as strong as $\sim 50$\% of H$\alpha$, we relax the above criterion for identification of a SNR to the condition $F_{\rm F673N}/F_{\rm F657N} > 0.3$. Because of their very red colors, old clusters may sometimes masquerade as SNR candidates. However, these may be simply removed on two considerations -- first, they are centrally concentrated, while the SNR candidates appear as shells and second, unlike \emph{bona fide} SNR candidates, these clusters are very faint in the [\OII] filter.

Following identification, the diameters and the positions were measured. The accuracy of the coordinates was checked by comparing the positions of the twelve candidates which are in common with the catalog of \citet{BL04}. The agreement is excellent.

The resultant catalog of the 40 SNR candidates found in the disk is given in Table \ref{tab:SNRlist1}. The objects are named by galaxy, nature and field (\emph{viz.} M83-SNR-1-) and are listed in order of increasing RA. The more interesting characteristics of the candidates are given in the notes. Several of the candidates have already been identified as X-ray sources \citep{SoWu02,SoWu03,Kilgard05} or radio sources \citep{Maddox06}. Only one ( SN1968L) of the six historical SNe listed in \cite{BL04} lies within our field. This case is discussed in more detail below.

\begin{figure}
\includegraphics[width=\hsize]{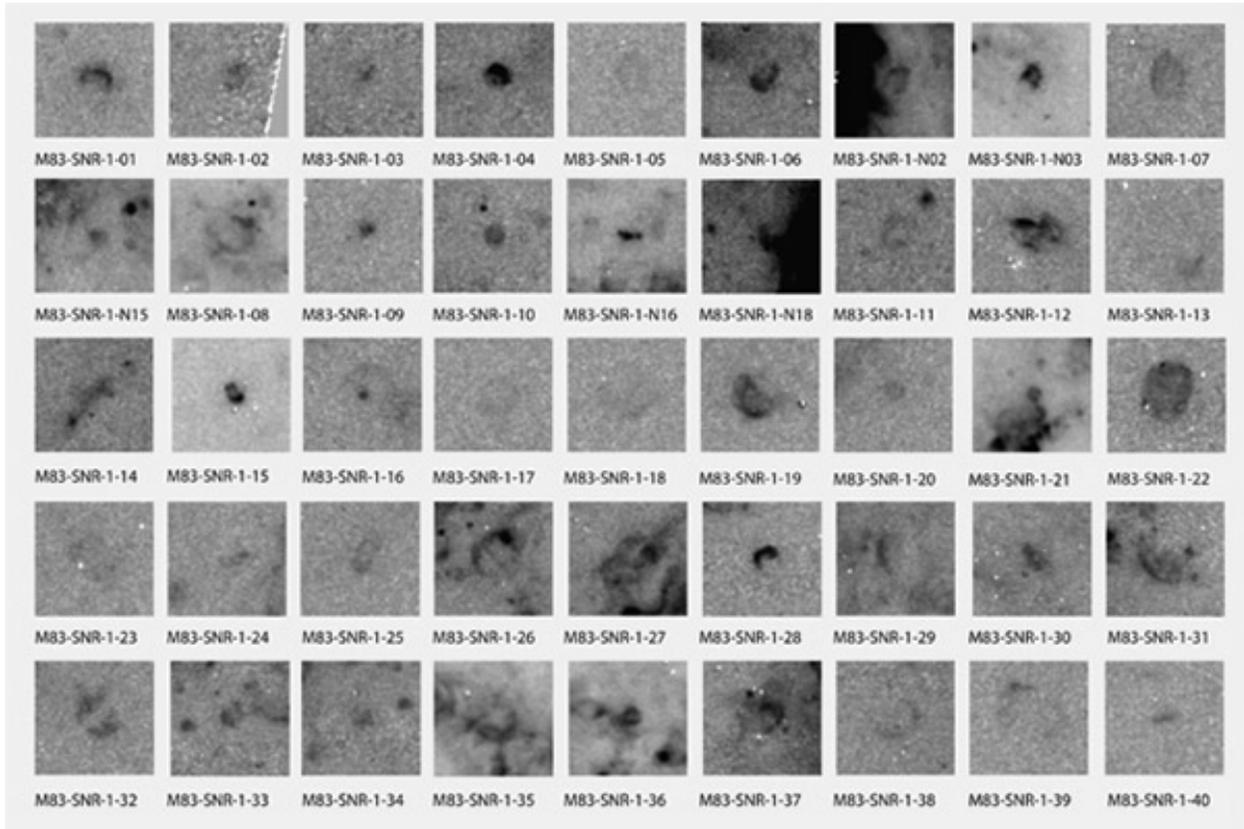}
\caption{Continuum subtracted H$\alpha$ images of the candidate SNRs found in the disk of M83. Each thumbnail is centered at the position of the candidate. The images have a square root stretch, and each covers $3\times3$~arcsec. on the sky. }\label{fig:M83_SNRs} 
\end{figure}

The H$\alpha$ images of the disk SNR candidates are shown in Fig~\ref{fig:M83_SNRs}. Here, the images are shown on a square root stretch to bring out the fainter features. In all cases these SNR candidates are clearly resolved, and details of their internal morphologies are evident. The sizes of the SNRs are rather similar because all of these candidates are seen in their radiative phase. To first order their expansion velocities should be quite similar - governed by the timescale for the shock ISM gas to produce a fully radiate shock, and the size will be moderated only by the density in the surrounding ISM and the explosion energy.

Note that, in our images, we find several \HII\ regions with large central cavities. None of these have a detected SNR lying in them. This supports the idea that the collective effects of stellar winds and ablation of un-ionized inclusions by photoionization has effectively swept these cavities clear of \HI\ so that radiative SNR cannot be observed because the cooling timescale of the hot, tenuous ISM is too long.

A few of these SNR candidates are worthy of special mention. Especially interesting is M83-SNR-1-14 \emph{alias} BL39, \citet{BL04}. This has a linear bipolar structure with two bright knots at either end of the long axis, and a more diffuse nebulosity in between, see Fig~\ref{fig:M83_SNRs}. Its structure is similar to a double-lobe radio source, and is reminiscent of the famous galactic supernova remnant W50 \citep{Zealey80} associated with the black hole binary SS433 \citep{Margon79a,Margon79b,Blund04,Blund05}. We may speculate that M83-SNR-1-14 too has a black hole binary at its core. BL 39 was one of the objects for which \citet{BL04} obtained spectra, but apart from the relatively weak [\OIII] emission, it is not remarkable in its spectral characteristics. Higher resolution spectra would be useful to establish whether there is any evidence for a bipolar outflow in this object.

One might expect that an SS433-like source should be detected as an X-ray source. However, M83-SNR-1-14 (BL39) was not detected with Chandra in X-rays by \citet{SoWu02,SoWu03} down to the brightness limit of their survey; $3\times10^{36}$~erg~ s$^{-1}$.   However, Namiki et al. (2003) give the ``typical" 1-10 keV flux of SS433 as $8.7\times10^{-11}$~erg~ cm$^{-2}$~s$^{-1}$ which would correspond to a luminosity of $2.4\times10^{35}$~erg~ s$^{-1}$ at the distance of M83.  Thus, an SS433-like source would not necessarily be detected in X-rays with the current sensitivity.

A second SNR with a decidedly bipolar morphology is M83-SNR-1-12 (BL37). This object  also has spectoscopic data, and has strong [\OIII] and [\OII] emission.

\clearpage
\clearpage

\subsection{Nuclear Candidates}
Identification of the SNR in the nuclear region is rather more difficult for several reasons. First, there is extensive diffuse interstellar emission which is also characterized by a high [\SII]/H$\alpha$ ratio. Second, the extreme reddening in some regions, coupled with the older underlying stellar population introduces an additional color term in the star subtraction process which leaves a residual in the [\SII] frame. Third, the starburst nature of the nucleus produces very strong and finely structured H$\alpha$ emission, which makes the contrast between the \HII\ regions and the SNR shells very low. 

As a consequence of these problems we used a different technique in the nucleus. Judging from the weakness of [\OIII] $\lambda 5007$ emission in the whole of this field, we may safely assume \citep{Kewley02} that the chemical abundances in the ISM are above, and possibly much above, the solar values. With such high chemical abundances, the temperatures of \HII\ regions are quite low, possibly below 5000K, and at these temperatures collisional excitation of the [\OII]$\lambda\lambda$ 3727,9 emission lines is effectively suppressed. In shock-excited gas, however, the cooling region in which we find the \OII ion is characterized by a much higher temperature ($\sim 12000$K), and the collisional excitation rate of the [\OII]$\lambda\lambda 3727,3729$ emission lines is high. We therefore mainly rely on the strength of the \OII emission to identify SNR candidates in the nuclear region. Because of the high and variable dust extinction in the nuclear region, the [\OII] line is heavily attenuated. As a consequence, all the nuclear candidates that lie on the far side of the dust mid-plane will have been missed. This implies that our completeness in this region cannot be appreciably better than 50\%. The list of nuclear SNR candidates so identified is given in Table \ref{tab:SNRlist2}. These are listed by increasing RA and are named by galaxy, nature and field, and prefix (N) to identify them as nuclear candidates (\emph{viz.} M83-SNR-1-Nxx), For reference, we also give the measured coordinates of the nuclear cluster in this Table.

For those candidates which are not too affected by crowding with \HII\ regions, we have measured their  H$\alpha$ fluxes using the \textsf{digiphot} aperture photometry routines in IRAF. Because (at the time of writing) the on-orbit photometric zero point of the F657N filter had not been determined, the transformation between electrons per second and flux was determined by comparison with the fluxes given in the \citet{BL04} catalog for the dozen candidates we have in common. In some cases, we find \HII\ regions close to or within the boundaries of the SNR, and these objects were excluded from the fit.  We estimate that the resultant fluxes are good to about $\pm 25$\%. As in \citet{BL04} we make no attempt to correct the observed fluxes for the contamination by the [\NII]$\lambda 6584$  line. Accounting for this could reduce the measured flux by as much as $20-40$\%. However, an accurate determination of this correction will have to await spectrophotometric observations.

\begin{figure}
\includegraphics[width=\hsize]{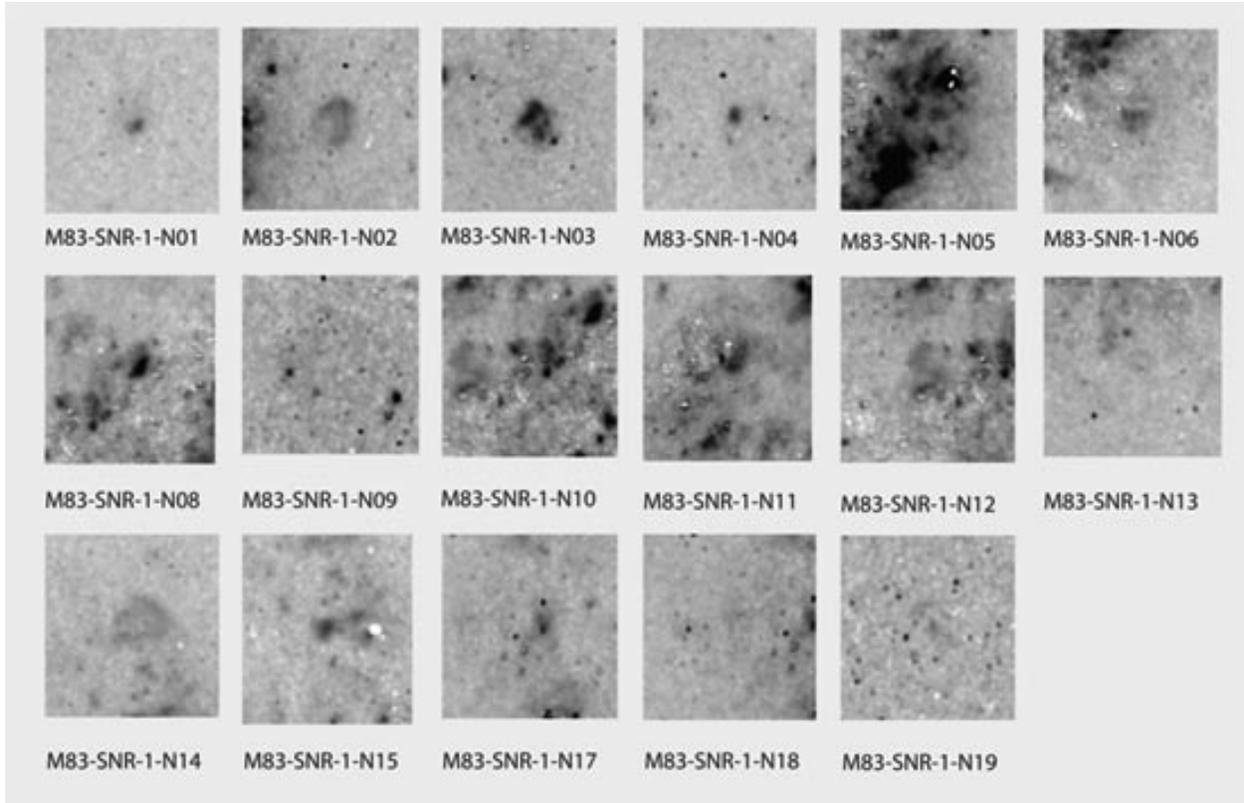}
\caption{Continuum subtracted [\OII] images of the nuclear candidate SNRs. Each thumbnail is centered at the position of the candidate, is on a square root stretch, and covers~$3\times3$~arcsec. on the sky. The small dark specks represent incompletely subtracted cosmic ray events, while the white specks are over-subtracted blue stars.\newline}\label{fig:M83_SNRs_Nuclear} 
\end{figure}

Images of the nuclear candidates are shown in [\OII]$\lambda\lambda 3727,9$ emission in Fig~\ref{fig:M83_SNRs_Nuclear}. Again, the images are shown on a square root stretch to bring out the fainter features. Each thumbnail covers~$3\times3$~arcsec. on the sky. In the nuclear candidates  the [\OII] emission is very strong compared to underlying \HII\ regions, and this provides an excellent contrast between the SNR and the background.  
\clearpage
\clearpage

\subsection{Extended O-strong nebulae}
As a result of the high chemical abundance in this field, [\OIII] emission is generally weak. Therefore, every object with significant [\OIII] emission is interesting in some way or another. The point sources with strong [\OIII]  are planetary nebulae (PNe), and we detect dozens of un-cataloged  PNe in this field. \citet{Herrmann08} have reported 241 PNe in M83, although the vast majority of these lie outside the WFC3 field. All five catalogued PNe that are in the WFC3 field have been detected as point sources in the [\OIII] frame.

Extended sources enhanced in [\OIII] may be either Wolf-Rayet ring nebulae, or else young supernova candidates in which we directly see the ejecta of the supernova event as it passes through the reverse shock. It is also possible that shocks of more normal SNRs may appear strong in [\OIII] emission for a brief time at the start of their radiative phase.

Fairly strong extended [\OIII] was detected from many of the SNRs listed by BL04 because shock velocities above 100 $\rm km ~ s^{-1}$ can still doubly ionize oxygen.  However, we also might expect any young, core collapse SNRs (similar to Cas A in our Galaxy or 1E0102-7219 in the Small Magellanic Cloud) to be dominated by [O~III] emission lines. Indeed, the emission lines normally used to find ISM-dominated SNRs may not be present at all.  

\begin{figure}
\begin{center}
\includegraphics[width=8cm]{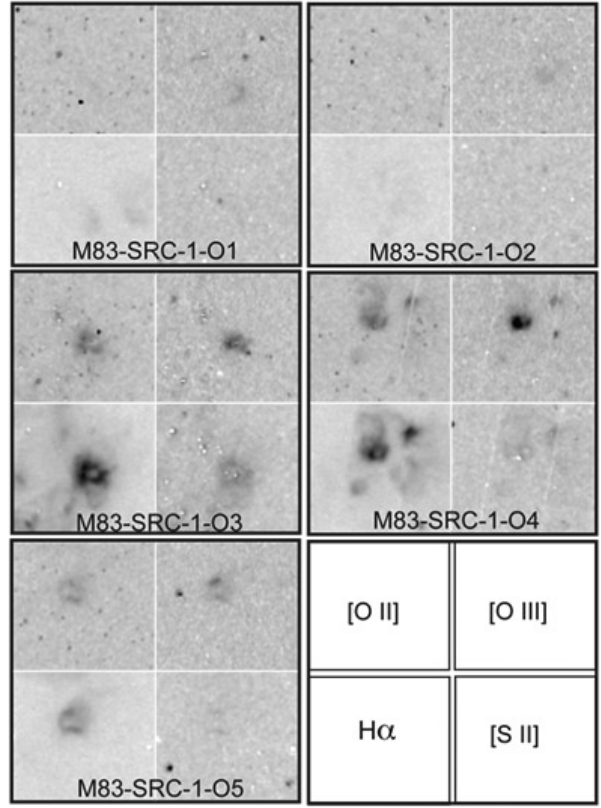}
\caption{Emission line images of the objects with unusually strong and extended [\OIII] $\lambda 5007$ emission. The key to the emission line corresponding to each image is given at the bottom right hand corner. The thumbnail images are each $3\times 3$ arcsec on the sky.  The lower three objects may either be O-rich young SNR or else high excitation Wolf-Rayet ring nebulae.}\label{fig:M83_O1} 
\end{center}
\end{figure}

For these reasons, we concentrated on inspecting any obviously extended sources of [\OIII] emission in the WFC3 field that did not already correspond with known or newly identified SNR candidates.  There are only a handful of reasonably bright [O~III] sources, the five most interesting of which are listed in Table \ref{tab:strongOlist}.  This may be a heterogeneous sampling of objects, and their true identity will require follow-up spectroscopy to decipher, but several of the objects appear consistent with a possible O-rich SNR interpretation.  The continuum subtracted emission line images of these sources are shown in Figure \ref{fig:M83_O1}. For each object there are four panels corresponding to their appearance in [\OII], [\OIII], H$\alpha$ and [\SII], respectively. We now discuss the properties of each of these sources in more detail.

\noindent{\bf M83-SRC-1-O1} is a small arc of unusually strong [\OIII] emission that aligns with a faint diffuse patch of H$\alpha$.  No enhancement in either [\SII] or [\OII] is seen and there are no strong stellar sources coincident with the region.The nature of this object remains unclear.  
\newline
{\bf M83-SRC-1-O2} is a second case of a (very) faint, extended [\OIII] nebula within a larger region of faint H$\alpha$ emission.  Its nature is unclear
\newline
{\bf M83-SRC-1-O3} is a clustering of O-knots centrally located within an H$\alpha$ and [\SII] shell which itself lies within a fainter, more extended nebulosity.  Coincident stellar sources complicate the interpretation somewhat, but this object looks like a possible analogue of N132D in the Large Magellanic Cloud, which $\sim3200$ years after the explosion has an outer $\sim$25 pc ISM-dominated shell surrounding inner O-rich debris from the SN \citep{Blair00}. If this analogy is correct, the outer shell in this case is roughly half as large as the one in N132D (0.36 $\times$ 0.48\arcsec, or 8.0 $\times$ 10.6 pc), which could indicate a smaller precursor-blown cavity surrounding the SN.
\newline
{\bf M83-SRC-1-O4} is a curious bright [\OII] and [\OIII] feature within a more extended region of H$\alpha$ and [\SII] emission.  It is by far the brightest object seen in the exposure using the WFC3 [\OIII] filter. There is significant H$\alpha$ emission coincident with the strong O-nebula, but [\SII] is weak.  The O-nebula appears as a partial shell open to the north, although a nearly coincident star or very tight cluster complicates the interpretation. In the V-band image the image is slightly fuzzy, which could be the result of the [\OIII] emission being passed by the broad V-filter.  This object could be either an O-rich SNR or possibly a high-excitation WR ring nebula.
\newline
{\bf M83-SRC-1-O5} is an interesting, relatively bright extended object. In H$\alpha$, a half shell is clearly seen open to the west and brightest on the north and south sides, with a diameter of $\sim$11 pc. Diffuse H$\alpha$ emission fills the shell and spills out to the west and north. The full ring is more obvious in [\OII], and is present but very faint in [\SII] (certainly too faint to be indicative of shock activity). The northern and southern H$\alpha$ bright spots are enhanced in [\OIII] (especially in the north) although the rest of the shell can be seen.  A bright continuum source is coincident not with the center of this shell but rather the northern edge, directly adjacent to the brightest [O~III] emission.  This nebula is probably photoionized and perhaps a WR shell of some type, but it is clearly unusual due to the strength of the [\OIII] line compared with most emission regions in the field.

Clearly follow-on spectroscopy is warranted for all of these interesting candidates. This is planned using the Wide Field Spectrograph (WiFeS) on the ANU 2.3m telescope \citep{Dop07}.
\clearpage
\clearpage

\subsection{Other X-ray / Emission line Sources}

\begin{figure}
\includegraphics[width=\hsize]{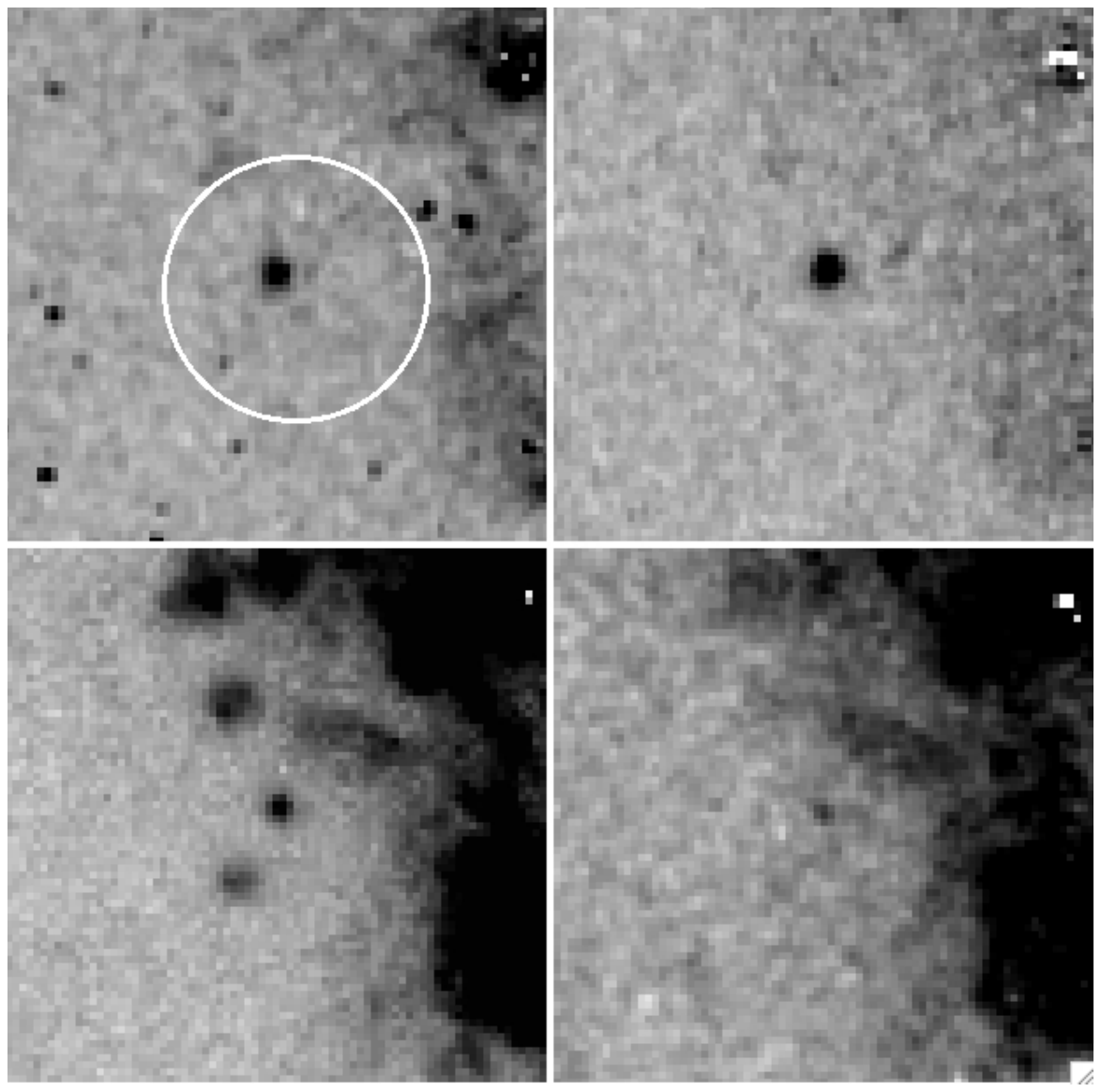}
\caption{This 4-panel figure shows a 3\arcsec\ region surrounding the position of Soria \& Wu (2003) Chandra source 70.  The white circle is 1.5\arcsec\ in diameter centered on the source 70 position.  Just northeast of the center is a nebular object with relatively strong [O~II] (upper left) and [O~III] (upper right) emission.  A compact H$\alpha$ source is also seen (lower left), but is much fainter in [S~II] (lower right).  This source is very likely a bona fide O-rich SNR in M83.}\label{fig:SW70} 
\end{figure}

In order to see whether any compact sources had been missed, we inspected the region around each of the 23 Soria \& Wu (2003) Chandra sources in the field that did not correspond to known or new SNRs found with the H$\alpha$ - [\SII] - [\OII] technique.  The most complicated regions of the nucleus were excluded from this search because of field crowding and source confusion.  From this exercise, one source stood out as having a high probability of being an O-rich SNR. This is  Soria \& Wu (2003) source 70, located just to the east of the bright nuclear starburst region.  Figure \ref{fig:SW70} shows a 3\arcsec\ region near this source (same panels as used above), where the white circle (1.5\arcsec\ diameter) is centered on the nominal Soria \& Wu position; RA (J2000)= 13:37:01.28, Dec. (J2000)=  -29:51:59.9.  A relatively strong and compact nebula is found in both [\OII] and [\OIII], slightly northeast of the nominal position. A slightly diffuse H$\alpha$ nebula also visible at this position, but the object is very weak in [\SII].  Although the V-band is not shown, there is no appreciable continuum source at this position, and so a stellar residual or stellar confusion is not an issue.

Table A.1 of Soria \& Wu (2003) indicates that source 70 is a soft X-ray source, with roughly 70 counts in the 0.3 - 1 keV band and 40 counts in the 1 - 2 keV band. This is consistent with expectations for a young, Cas A-like SNR.  We identify the  nebular object in Figure \ref{fig:SW70} as the likely optical counterpart of the X-ray source, and  a bona fide O-rich SNR in M83. The faint H$\alpha$ emission might arise in the X-ray ionized nebula surrounding the blast wave.
\clearpage
\clearpage

\subsection{The Search for the Remnant of SN1968L}
Of the six historical supernovae observed in M83 since 1923, only one, SN1968L, lies within our field of view. This event was observed in the western side of the nuclear region and was classified as a Type II, but its position is poorly known. Published positions are notoriously inaccurate (see discussion in BL04).  We have therefore reviewed the literature and reconstructed the position based on the images provided by \citet{Wood74}. Scaling the WFC3 F555W image to approximate the appearance of Wood \& Andrews Figure Ic, we can reconstruct the position of this object. We estimate this new position to be accurate to better than 1\arcsec\ and our derived coordinates should be good to the accuracy of the WCS solution used for the other SNR candidates.

\begin{figure}
\includegraphics[width=\hsize]{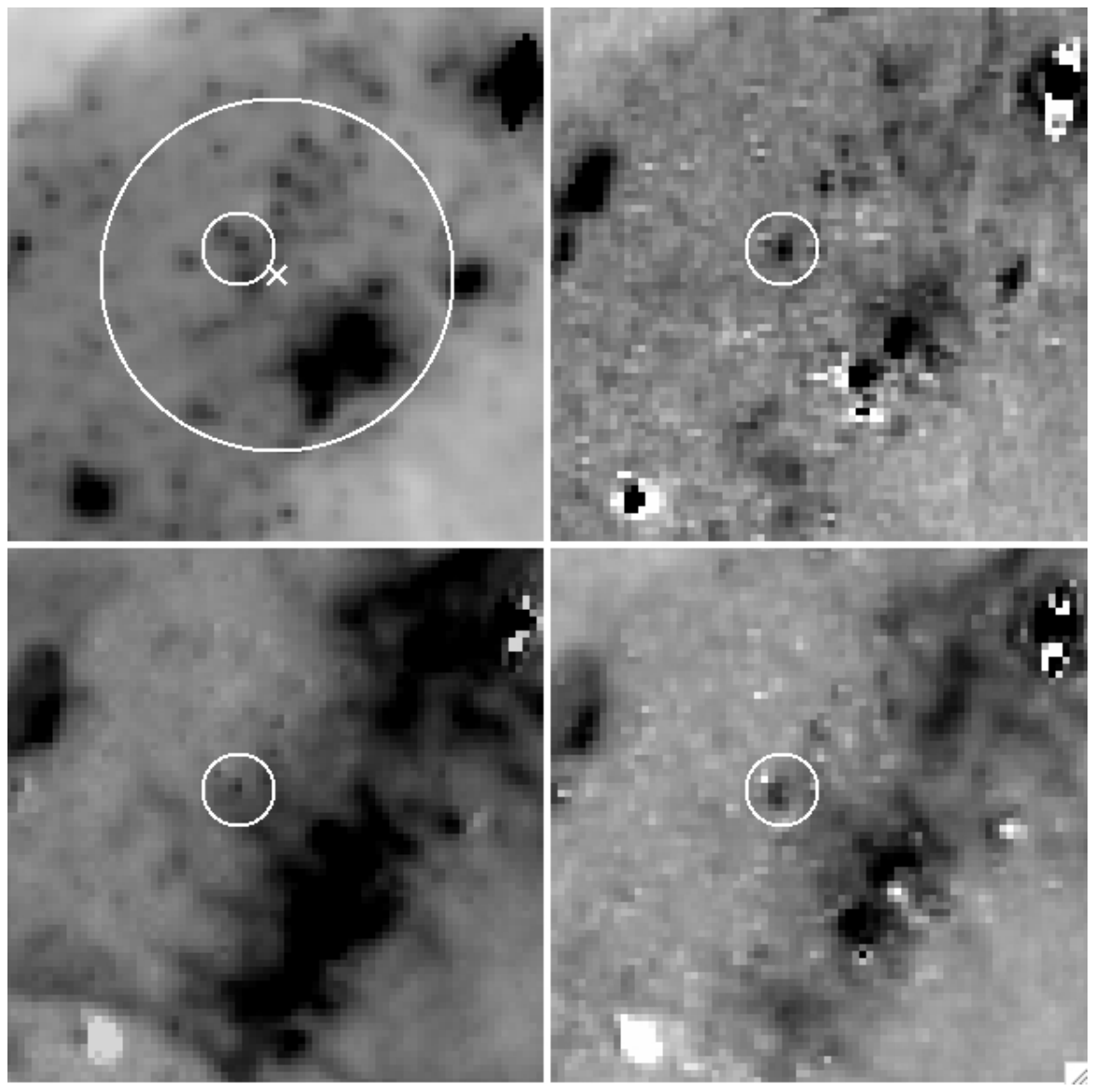}
\caption{A 3\arcsec\ field of WFC3 data centered on the expected position of SN1968L.  North is at the top, east at the left. The panels include (upper left) V-band, (upper right) [\OIII], (lower left) H$\alpha$, and (lower right) [\SII].  All images are shown with a square root scaling and all emission line images have been continuum-subtracted.  The X marks the reconstructed SN1968L position as described in the text, and the large circle (1\arcsec\ radius) is the conservative upper limit on its accuracy.  The smaller (0.2\arcsec\ radius) circle marks the optical counterpart candidate for the SN, based on the appearance in the [\OIII] image.  This circle is reproduced in the other panels for reference.  A faint stellar source lies just NE of the [\OIII] source.}\label{fig:SN1968L} 
\end{figure}

This region is marked on appropriately scaled WFC3 images of the nuclear region, \emph{see} Figure \ref{fig:SN1968L}.  Inspection of these images reveals an emission nebula with a strong, probably unresolved core and a fainter, marginally resolved halo in [O~III]. This has a coincident slightly extended [S~II] nebula at about 0.2~arcsec. distant from our reconstructed position of the supernova event.  No obvious H$\alpha$ counterpart is present although the complexity of H$\alpha$ emission and the faintness of the object against the strong H$\alpha$  background could mask this.  No obvious counterpart is present in the [\OII] waveband, but this could simply be due to heavy foreground dust attenuation.  The reconstructed position for this object is RA 13:37:00.433 and Dec. $-$29:51:59.65 (J2000).

What might be expected for such a young SNR? If we assume the ejecta have been expanding at $\rm 10^{4} ~ km ~ s^{-1}$, it would still be less than 0.5 pc in diameter, which corresponds to $<$0.022\arcsec.  The unresolved nature of the core [\OIII] source is thus consistent with a possible young supernova remnant identification.  It is not entirely clear whether the [\SII] nebula is extended or whether the appearance is impacted by the continuum subtraction of nearby objects.  We note, however, the similarity to Cassiopia A in our Galaxy where oxygen and sulfur dominate the appearance of the optical spectrum.  The slightly extended halo in [\SII], and the corresponding [\OIII] - emitting region may possibly arise from blast-wave photoionized pre-supernova stellar ejecta, similar to that seen in SN1987A \citep{Fran89,Crotts91}. If placed at the distance of M83, this ring-like nebulosity would have a diameter of $\sim0.06$\arcsec, and be marginally resolved on these WFC3 images.

In conclusion, we tentatively claim the object in the small  circle in the Figure to be the recovered optical counterpart to SN1968L.  A spectrum showing high [\OIII] velocities would confirm this identification.

\section{Interpretation of the Observations}\label{sect:Interpret}
\subsection{Cumulative Number : Diameter Relationships}
According to classical theory, a SNR evolves through a number of different phases in its passage to the point where it merges with the surrounding interstellar medium. These are summarized in \citet{DopSuth03}. In brief, the SNR first expands ballistically, $R\propto \tau$ until it has swept up sufficient mass to allow a reverse shock to propagate through the ejecta to heat it to very high temperature. The SNR then expands adiabatically in its Sedov phase $R\propto \tau^{2/5}$ until the outer blast wave shock becomes slow enough and old enough to start radiating away the energy stored in the hot gas. During the ballistic, reverse shock and Sedov phases the SNR is seen mostly at X-ray or radio frequencies, except for a short period when stellar ejecta passing through the reverse shock becomes prominent at optical wavelengths, especially in the lines of [\OI], [\OII], [\OIII] and [\NeIII].  In the following, radiative, phase $R\propto \tau^{2/7}$ \citep{McKee77}. This is the phase when the SNR is most likely to be observed at optical wavelengths, and in this phase the abundances inferred from the radiative shock theory will reflect those of the surrounding ISM rather than the stellar ejecta. Finally, the energy contained in the hot plasma is radiated away, and the SNR enters its final momentum-conserving  ``snowplough" phase \citep{Oort46} with  $R\propto t^{1/4}$ before it turbulently merges with the general ISM.

Since the M83 SNR are detected at optical wavelengths, they are either in the radiative phase, or else in the (ejecta dominated) reverse shock phase. In our [\SII] and [\OII] selected sample, none appear to be young ejecta-dominated SNR. Therefore to the degree that we can regard supernova as exploding at a constant stochastic rate, we would expect the cumulative number : diameter relation to follow $N(<D) \propto D^{7/2}$. 

\begin{figure}
\includegraphics[width=\hsize]{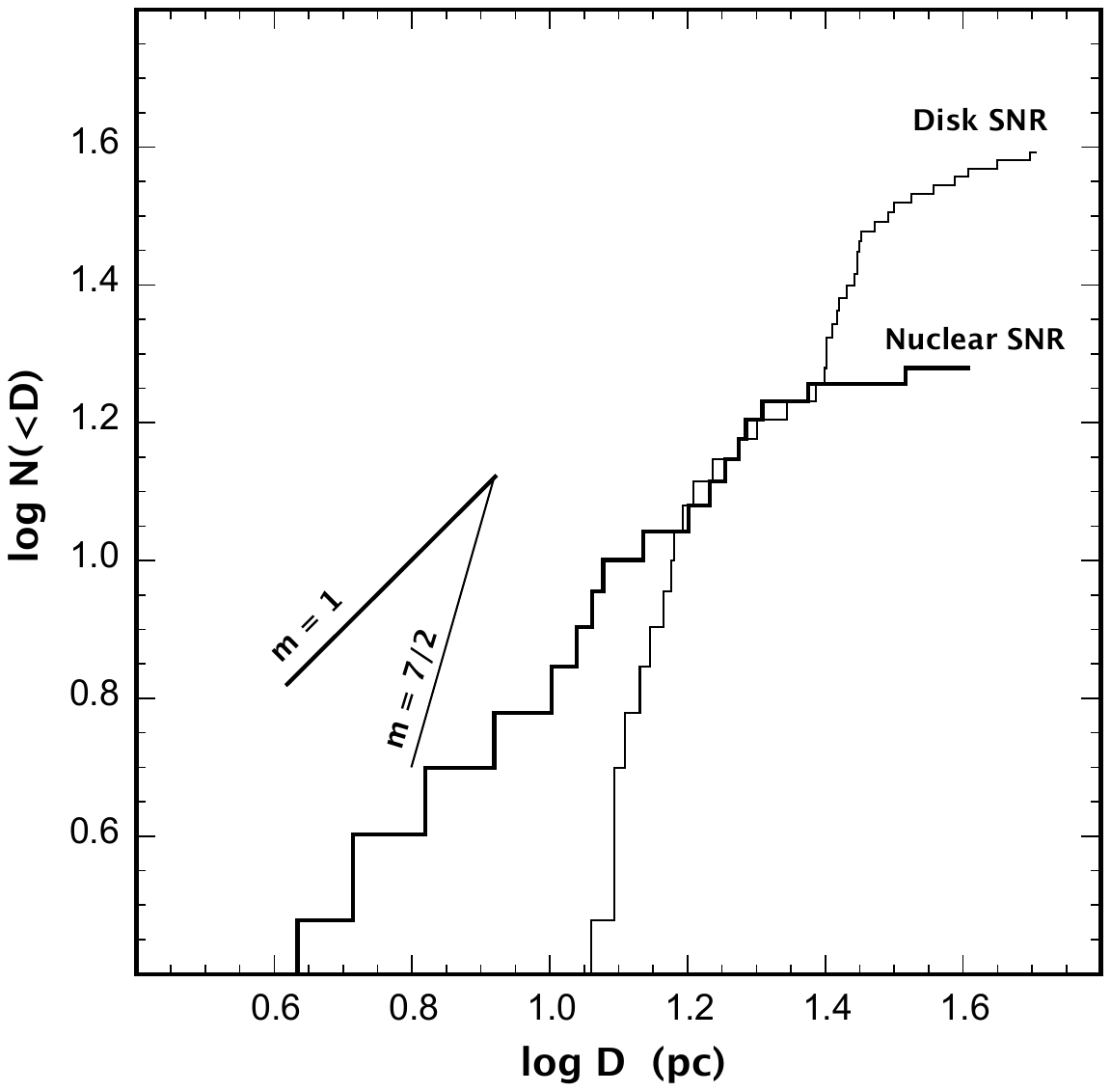}
\caption{The cumulative number: diameter relation for the M83 SNR. Here we distinguish the disk SNR population from the SNR discovered in the nuclear regions. The slope, $m$, of the relation for the  disk SNR is consistent with these SNR being in their radiative phase ($m=7/2$). The fluctuations around the slope reflect the stochastic timing of the supernova events. The nuclear SNR follow a slope which suggests that they are mostly in free expansion ($m=1$) until the blast wave collides with a dense cloud in the ISM, propagating a radiative shock into it. \newline}\label{fig:1} 
\end{figure}

The observed distribution is shown in Figure \ref{fig:1}. The SNR divide according to their environment - the nuclear group is consistent with being observed in free expansion, while the disk SNR generally follow the slope expected for SNR in their radiative phase. Superimposed on this slope are clear stochastic fluctuations due to the randomness of SNe. 

Why should the nuclear population be distinct from the SNR in the spiral arms? The most likely explanation for this is that there is a high porosity in the nuclear ISM. Given that it is a high-pressure region, the un-ionized part of the ISM is likely to be crushed to higher density, and the filling factor of the hot ISM is correspondingly larger. In sweeping up the hot tenuous ISM, the SNR blast wave would not be appreciably slowed until it encounters one of the dense \HI\ clouds, driving a much slower radiative shock into it. At this point, it becomes very bright in optical emission lines. According to this scenario, the pre-shock density in the nuclear SNR should be much higher than in the disk SNR. We will show below that this is indeed the case.

For the nuclear SNR we can estimate both the ages and the supernova rate in the nucleus by assuming that in its free expansion phase the SNR expands at 5000-10000 km~s$^{-1}$. There are 11 SNR with a diameter less than 15~pc, with an age $<750-1500$~yr. This implies a supernova rate of one every 70-140~yr. Incompleteness in our SNR catalog may almost double this estimated rate to one every $\sim50$~yr. These estimates are consistent with the fact that one SN has been observed in this region since 1923.

\subsection{Derived SNR Parameters}\label{Parms}

With the H$\alpha$ photometry and the measured diameters, we can derive some interesting parameters for these SNR candidates. These parameters are summarized in Table \ref{tab:SNRparms}. 

The total luminosity, $L$, can be derived directly from the H$\alpha$ luminosity. For this, we use the distance given above and use the transformation $L=75L_{ \rm H\alpha}$ (for a shock velocity of 200 km~s$^{-1}$. This transformation is derived from the shock models of \citet{DopSuth96} and \citet{Allen08}. In practice, the relationship between the H$\alpha$ luminosity and the total luminosity is weakly dependent on the shock velocity. Between 200 km~s$^{-1}$ and 500 km~s$^{-1}$, the total luminosity relative to the H$\alpha$ luminosity increases by a factor of two. 

For the M83 SNR we adopt a characteristic shock velocity, $v_s$, of 200 km~s$^{-1}$. This figure is chosen on the basis of that it is typical of the optical velocity dispersions observed in bright radiative supernova remnants in our Galaxy (\emph{e.g.} \citet{Shull91}) and in the Magellanic Clouds \citep{Mea87,Shull83,Bil07}. An upper limit can be derived from the fact that the [\NII] and H$\alpha$ lines are clearly resolved in the low resolution spectroscopy of the M83 SNR by \citet{BL04}. This limits the SNR expansion velocities to less than $\sim 300$~km~s$^{-1}$. An extreme upper limit on the shock velocity is placed by the condition that the cooling timescale for the shocked gas has to be shorter than the dynamical age of the SNR - otherwise the shocks would only be visible in X-rays. The cooling timescale for a radiative shock is given by \cite{DopSuth03}:
\begin{equation}
\tau _{\rm cool}\sim 200\frac{{\rm v_{100}}^{4.4}}{Zn_{0}}\,\mbox{yr},\label{eqn:tcool}
\end{equation}
where ${\rm v}_{100}$ is the shock velocity in units of 100 km~s$^{-1}$, $Z$ is the chemical abundance in the pre-shock gas relative to solar, and $n_0$ is the pre-shock number density. In what follows, we will take $Z=1$. In practice,$Z$ is probably rather higher than solar. All of our SNR candidates are, by definition, in their radiative phase. Thus, if we can estimate $n_0$, we can then use equation \ref{eqn:tcool} in conjunction with the dynamical age to estimate the maximum shock velocity consistent with the SNR being in its radiative phase. 

We may use the total luminosity to estimate the radiative lifetime of these SNR. The radiative lifetime is defined by:
\begin{equation}
\tau _{\rm rad}= \frac{E_0}{L}\,\mbox{sec},\label{eqn:trad}
\end{equation}
where $E_0$ is the initial kinetic energy of the SNR ejecta. Here we take $E_0 \sim 10^{51}$~ergs. We now use the area of the SNR shock defined by the mean optical diameter, as measured from the SNR image to estimate the mean surface brightness of the shock wave, $S$ (erg cm$^{-2}$ s$^{-1}$).

This radiative lifetime simply tells us the maximum time over which it can remain in the radiative phase before the stored explosion energy is lost as readiation. The radiative lifetime is correlated with the true age of the SNR, but not in any simple manner. Theoretically, the SNR has already passed through its free expansion and Sedov (adiabatic) phases before it reaches the radiative phase, so the actual age of the SNR depends on how long it has spent in each of these, and also on how long it has already spent in the radiative phase before being observed.

Next, we use this surface brightness to estimate the pre-shock density, $n_0$, derived from radiative shock theory \citep{DopSuth96,Allen08}. For a fully radiative shock, the surface energy flux across the shock front is, $S=\rho v^3 /2$. Transposing to extract the number density, and assuming a Helium abundance by number of 0.1, this equation can be written as:
\begin{equation}
\log\left({n_0/ {\rm cm^{-3}}}\right) = 2.88 + \log S -3\log {\rm v}_{100}. \label{eqn:density}
\end{equation}
Note that the pre-shock density derived in this way is highly sensitive to the assumed shock velocity. Clearly, a measurement of the velocity dispersions of thes SNR would considerably assist in the accuracy of the derivation of the inferred parameters for individual SNRs. The pre-shock densities so derived lie in the range $0.6 < n_0/cm^{-3} < 1700$ (see Table \ref{tab:SNRparms}). These densities are consistent with the range of volume densities inferred by \citet{Heiner08} at the surface of the giant molecular clouds in M83; $0.1-400$ cm$^{-3}$ for the region of M83 lying within $R_{25}$. Since the phase of the ISM which these densities trace is probably that of the warm neutral medium (WNM), the appropriate temperature is probably $\sim 5-10\times10^{3}$~K, which implies that the range of ISM pressures in M83 is of the order $5\times10^{3} < P/k < 1.5\times10^{6}$~cm$^{-3}$K.

Finally, as described above we use the density derived from equation \ref{eqn:density} in conjunction with the radiative age given with equation \ref{eqn:trad} and the cooling timescale given in equation \ref{eqn:tcool} to estimate the \emph{maximum} shock velocity consistent with the requirement that the SNR is in the radiative phase. The values so determined are listed in Table \ref{tab:SNRparms}. As expected, all of these are larger than the 200  km~s$^{-1}$ adopted as the likely shock velocity

\subsection{The Radiative Age : Density Relationship}\label{sec:den-rad}
\begin{figure}
\includegraphics[width=\hsize]{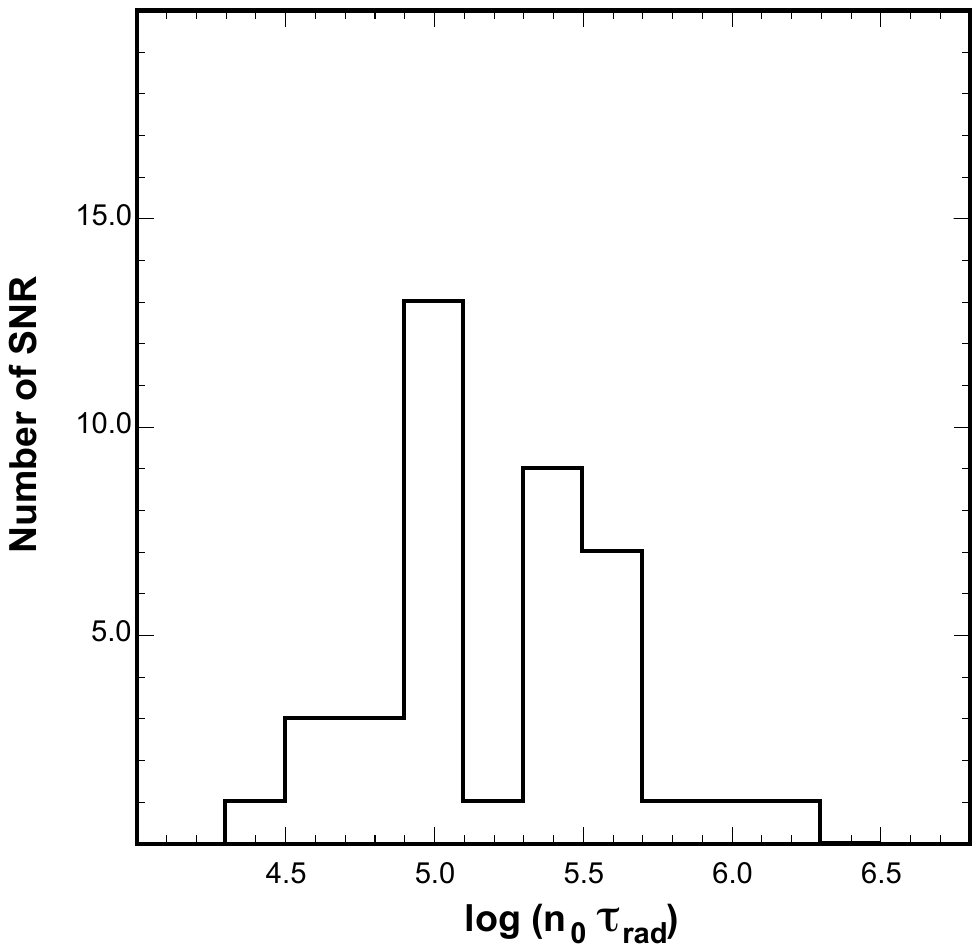}
\caption{The histogram of the product $n_0\tau _{\rm cool}$ for our SNR candidates.  The width of this should reflect both the range of local physical conditions in the ISM and the range of explosion energies of the objects. Remarkably, the histogram is clearly bipolar about $\log (n_0\tau _{\rm cool}) = 5.2$, with the two peaks being separated by 0.5-0.6 in the logarithm of the product. \newline}\label{fig:Histogram} 
\end{figure}

Because the cooling timescale of the shocked gas scales inversely as the pre-shock density, the distribution of the product $n_0\tau _{\rm cool}$ should trace out the degree either to which local conditions influence the evolution (through variation in local density), or alternatively, the range in the explosion energy of the SNR candidates, $E_0$ (which influences the radiative cooling timescale). The histogram of  the product $n_0\tau _{\rm cool}$ is given in Figure \ref{fig:Histogram}. This distribution is clearly bimodal, suggesting that either the explosion energy is bimodal, or that there are strong and distinct environmental effects which form two populations of SNR.

\begin{figure}
\includegraphics[width=\hsize]{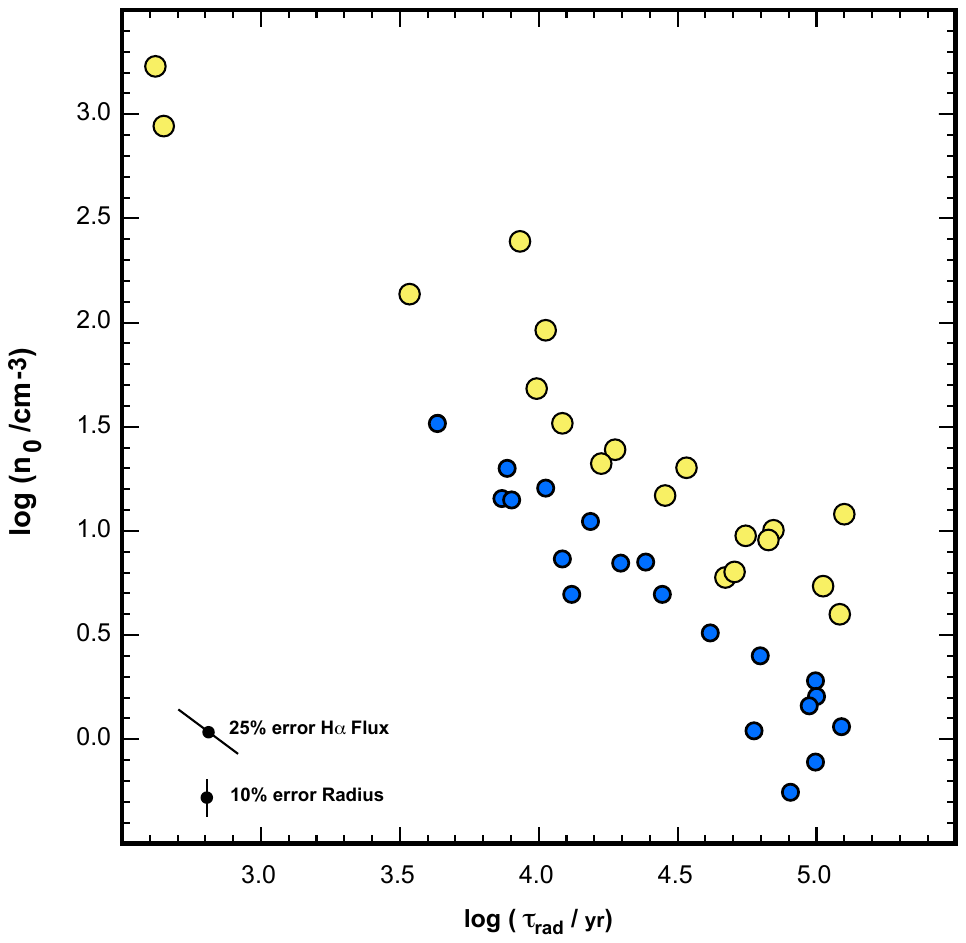}
\caption{The relationship between the radiative age and the derived pre-shock density of the M83 SNRs. The error bars represent the displacement that would be produced by a $\pm 25$\% error in the measurement of the H$\alpha$ flux, and by a $\pm 10$\% error in measurement of the SNR radius. The theoretical slope of this relation is -1. It is clear that there are two distinct populations of SNR distinguished here by the different symbols. For a given dynamical age the upper (yellow points, high density) sequence has a pre-shock density four times higher than the lower  (blue points, low density) sequence.}\label{fig:Age-Dens} 
\end{figure}

This bimodality can be investigated in a different way. We would also expect to see these SNR trace out a sequence with a slope of -1 on a $\log \tau _{\rm cool}$ : $\log n_0$ diagram. This is plotted in Figure~\ref{fig:Age-Dens}. Here we have divided the two populations at $\log \left[(n_0/{\rm cm^{-3}})(\tau _{\rm cool}/{\rm yr})\right] = 5.2$ into a ``high'' and a ``low'' sample distinguished by yellow circles and the blue circles, respectively. These two distinct sequences are separated by a factor of four in density at constant age, or else a similar factor in terms of age at constant density. These two groups of SNR are not identical with the nuclear and the disk groups defined above, although most of the nuclear objects fall within the ``high'' sample.

The likely measurement errors for the observable quantities which go into this relationship are shown as error bars in Figure ~\ref{fig:Age-Dens}. It is clear that separation between these two sequences is too large to be explained by measurement errors. It therefore probably represents either that there are two precursor populations of Type II supernovae, giving explosion energies separated by a factor four, or that there is an environmental difference which provides a density contrast of about the same factor.

\begin{figure}
\includegraphics[width=\hsize]{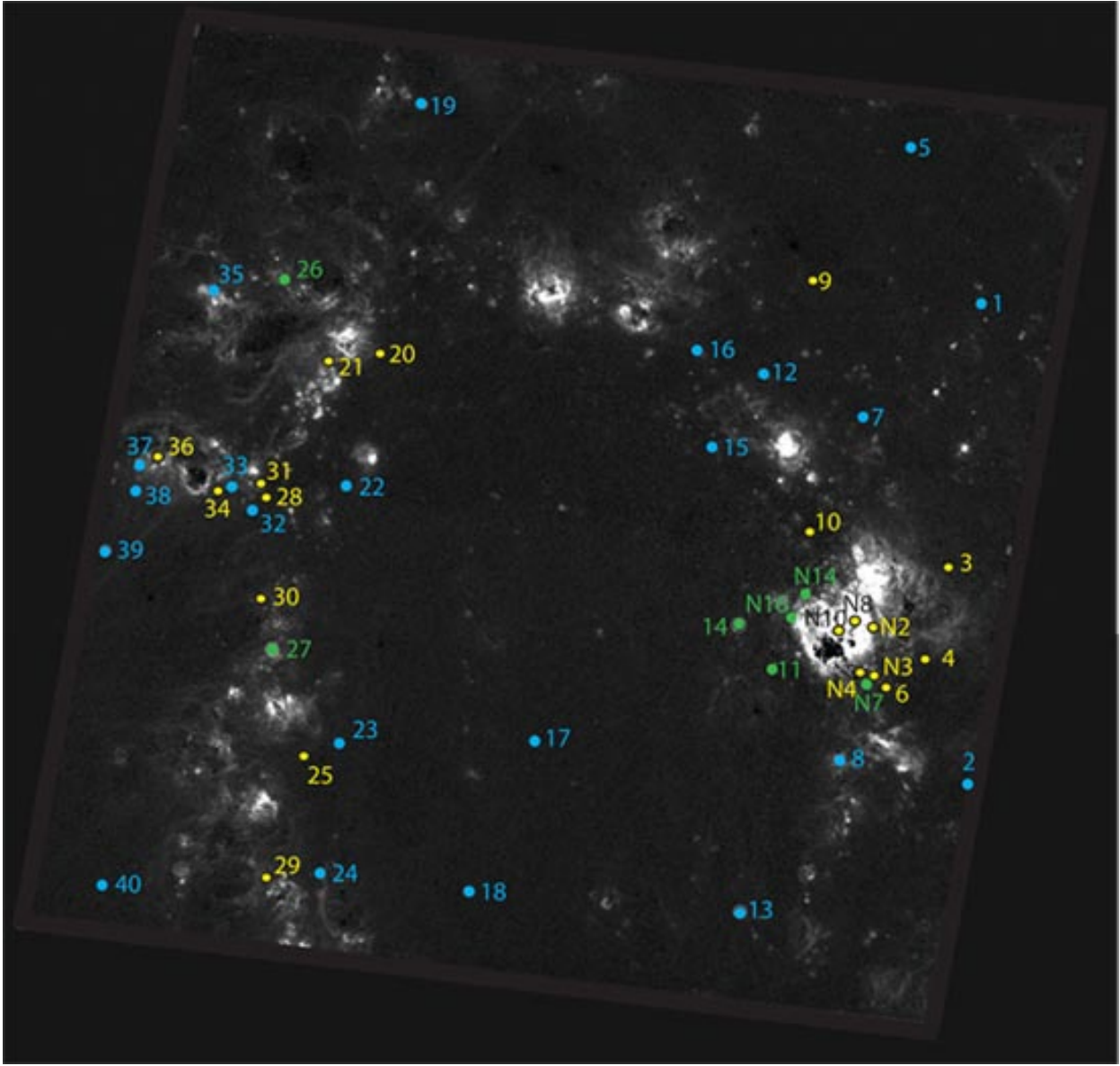}
\caption{The positions of the M83 SNR are here plotted on the H$\alpha$ image of M83. A number of the nuclear SNR are not shown because of crowding, and a lack of accurate photometry. Here we distinguish the objects by color. The yellow points represent the objects identified from figure \ref{fig:Age-Dens} as belonging to the ``high density'' sequence, the blue points represent the SNR in the ``low density'' sequence, and the green points are for unclassified SNR. It is clear that the high density sequence objects are all confined either to the nuclear region, or to the spiral arm.\newline}\label{fig:Atlas} 
\end{figure}

In order to test the environmental hypothesis, we plot the SNR on the H$\alpha$ image, distinguishing the ``high''  sequence points (yellow) from the ``low''  sequence ( marked in blue) - see Figure \ref{fig:Atlas}. For some of these SNR in locally crowded fields, it was not possible to measure the H$\alpha$ fluxes. These are shown in green. From this figure it is evident that the high density sequence objects are nearly all confined either to the nuclear region, or to the spiral arm, while the low density sequence is more generally distributed in the disk. The most likely explanation for this extraordinary result is that the environment of the spiral arms and the nucleus provides higher density \HI\ clouds for the SNR to interact with, most likely the result of a higher pressure in the interstellar medium in these environments.

Since the sound speed in the X-ray emitting plasma is very high, of order 300-500 km~s$^{-1}$, and the cooling timescale of this plasma is long, a local hydrostatic equilibrium can be achieved in this phase of the ISM. Therefore, in a medium with high porosity, the hot ISM can be thought of as supplying the substrate pressure confinement for the warm and molecular phases of the ISM. The local pressure in the X-ray gas is in turn correlated with the surface brightness, provided that the scale height of the X-ray emitting plasma and the temperature of this plasma do not vary too much across the face of the galaxy. Thus, the surface brightness in X-rays should correlate with the local density, $S_{\rm X-ray} \propto n_0^{2}$. 

We have measured the surface brightness of the X-ray emission using the Chandra observations described by \citet{SoWu03}.   These show diffuse emission concentrated toward the spiral arms and nucleus (in addition to the 127 point sources).  The diffuse emission arises primarily from hot gas, as evidenced by the fact that spectra show strong features from lines expected in plasmas of 10$^{6}$ - 10$^{7}$ K.  To determine the X-ray surface brightness at the position of each SNR, we created 0.4-2 keV images of the emission from M83, subtracting contributions from  foreground diffuse X-ray emission and an extragalactic background. We measured the average surface brightness within 5\arcsec\ of each SNR, omitting portions of the circular regions affected by point sources.  

\begin{figure}
\includegraphics[width=\hsize]{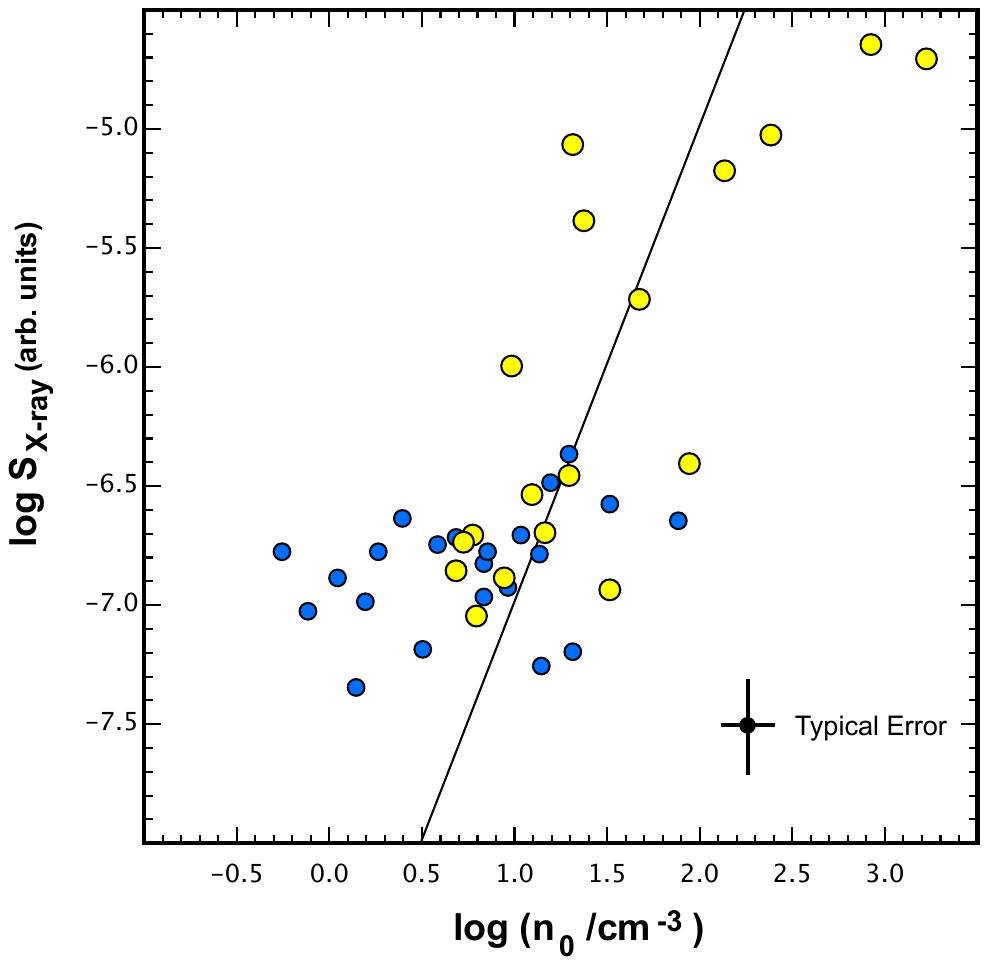}
\caption{The correlation between the derived pre-shock density of the M83 SNRs and the local (5 \arcsec\ radius) diffuse X-ray surface brightness. Scatter in this relationship is produced by variation in  X-ray scale height, X-ray temperature, absorption of the soft X-rays and extinction in the SNRs. The mean measurement error in the points is shown. If the X-ray plasma provides the substrate pressure in the ISM, and if scale height and the temperature of the X-ray plasma remained constant, then $S_{\rm X-ray} \propto n_0^2$.  A line of this slope is plotted on the figure. It is clear that only the ``high density'' sequence (yellow points) conforms to such a relationship. Other factors are determining the pressure in the ISM for the SNR in the ``low density'' sequence (blue points).}\label{fig:X-n0} 
\end{figure}

We compare the density derived from the optical data to the surface brightness (in arbitrary units) derived for the diffuse X-ray emission in  Fig \ref{fig:X-n0}. The line drawn on this figure has the theoretical slope  $S_{\rm X-ray} \propto n_0^{2}$. Scatter in this relationship is produced by variation in  X-ray scale height, X-ray temperature, absorption of the soft X-rays as well as all the factors which enter into the determination of the pre-shock density in the SNRs. The absorption of soft X-ray emission in the ISM is more likely to be important in the nuclear regions, where the ISM column density may be $\geq 10^{21}$ cm$^{-2}$. This affects the points with $n_0 > 500$ cm$^{-3}$. 

It is clear that there also is a bimodal behavior  of the points plotted on Fig \ref{fig:X-n0}. The ``high'' sequence points conform to the theoretically expected  $S_{\rm X-ray} \propto n_0^{2}$ relationship, showing that the X-ray plasma does indeed supply the substrate pressure to the ISM in both the spiral arms and in the nucleus. The ``low'' sequence shows no such relationship. For these objects, the substrate pressure in the ISM must be provided by some other source, such as interstellar turbulence. The diffuse X-ray emission which is spatially coincident with these SNRs probably originates in the halo of the galaxy, with a much higher scale height than in the arm regions, and is not directly physically connected with conditions in the disk. The virtue of Fig \ref{fig:X-n0} is that, unlike Fig \ref{fig:Age-Dens}, the bimodality cannot be ascribed to a bimodal behavior of an intrinsic property of the SNR population such as explosion energy $E_0$, but instead has to be related to the environment.

\subsection{Mass of Precursor Stars}\label{sec:mass}

The mean mass and lower mass limit of the SNe precursor stars can be estimated from the number of observed SNR - as long as we know both the star formation rate and the mean age of the SNR at the completeness limit in diameter. The argument goes as follows. 

First, we can use the total flux in H$\alpha$ in the field to estimate the total star formation rate. The theory behind this is that the flux in any hydrogen line is simply proportional to the number of photons produced by the hot stars, which is in turn proportional to the birthrate of massive stars. This relationship has  been well calibrated at solar metallicity for the H$\alpha $ line. In units  of M$_{\odot}$~yr$^{-1}$, the estimated star formation rate is given by \citep{Dopita94,Kennicutt98,Panuzzo03}:
\begin{equation}
\left[SFR_{{\rm H\alpha }} / \rm {M_{\odot}~yr^{-1}}\right]
=(7.0-7.9)\times 10^{-42}\left[ L_{{\rm H\alpha }} / {\rm erg.s}^{-1}\right] .  \label{eqn:SFR}
\end{equation}

For the full field of the continuum-subtracted WFC3 H$\alpha$ image, we measure a total luminosity of $L_{\rm H\alpha} =3.73\times10^{41}$erg s$^{-1}$, which translates by equation \ref{eqn:SFR} to a star formation rate of 2.76 M$_{\odot}$~yr$^{-1}$. This estimate does not take account of those \HII\ regions which are obscured by dust, or of the unknown contribution of the [\NII] emission within the bandpass of the filter.

The total number of SNR found in this image is 60. We may hope that any incompleteness in this number is matched by any incompleteness in our estimate of star formation rate, since both SNR and \HII\ regions are observed at H$\alpha$ and therefore subject to the same attenuation by dust. This assumption is certainly an over-simplification, but it cannot be tested in any simple way with our data.

In order to compare the expected number of SNR with the observed number we need to know both the shape of the upper initial mass function (which determined how many stars are available to explode as SNe for a given star formation rate), and the mean (physical) age of the SNR at the point where the sample becomes incomplete. From the cumulative number : diameter relationship, it is apparent that the sample becomes seriously incomplete at a diameter of $\sim30$pc at which point we have accumulated 46 SNR. We therefore attempt to estimate the age of the $D<25$pc sample, which is the largest diameter at which we think the sample is complete. However, at this diameter we do not know the mean physical age, only the observed range of radiative lifetimes; $3.8 < \log(\tau_{\rm rad}/{\rm yr}) <5.0$. Assuming that the radiative lifetime is comparable with the expansion timescale at the time they are observed, we can estimate the true age of the SNR to be on average 2/7 of the observed radiative age. This (very crudely) puts the mean age of the SNR in the complete sample at $\log(\tau/{\rm yr}) = 3.9\pm 0.5$.

Once we have settled on the form of the IMF, we can demand that the number of SNRs we see up to the completeness limit matches the observed number of SNRs. The mean lifetime of the SNR population at the completeness limit provides an estimate of the lower mass limit of Type II SNe. The form of the IMF also provides the mean mass of SNe. We have adopted a \citet{M-S79} IMF. This has 10.2\% of the stars (by mass) with mass above 9 M$_{\odot}$ and 3.6\% above 20 M$_{\odot}$. On this basis, we estimate that the lower mass of Type II SNe in M83 is $M_{\rm min} = 16^{+7}_ {-5}$~M$_{\odot}$ and that the mean Type II precursor mass is $M_{\rm av} = 24^{+10}_{ -7}$~M$_{\odot}$.

Whilst these number are necessarily very approximate, we have demonstrated that the number of SNR detected in these images is consistent with expectations of stellar evolution theory and with the observed rate of star formation in the observed M83 field.

\section{Conclusions}
This WFC3 dataset has provided an rich source of physical insight into the nature of the supernova remnant population in M83, and their interactions with the interstellar medium. To summarize our results:
\begin{itemize}
\item{We have increased by a factor of five the number of SNR candidates in this field.}
\item{We have likely recovered the remnant of the historical supernova, SN1968L.}
\item{We discovered that the nuclear ($R<300$~pc) SNR are physically distinct from those in the disk, being characterized by free expansion in a hot ISM with both high filling factor and pressure. We infer a supernova rate of around one every $\sim50$~yr, consistent with the fact that one supernova event has been observed in this region since 1923.}
\item{The range in pre-shock density that we infer, $0.6 < n_0/{\rm cm^{-3}} < 1700$, is consistent with the \citet{Heiner08} estimate of the densities at the surface of the giant molecular clouds of M83, obtained using GALEX far-ultraviolet (FUV) fluxes, VLA images of the 21-cm HI column densities, and estimates of the local dust abundances.}
\item{There are two distinct populations of SNR divided at $\log \left[(n_0/{\rm cm^{-3}})(\tau _{\rm cool}/{\rm yr})\right] = 5.2$ into ``high density'' and  ``low density'' samples. The ``high'' sample is mostly confined to the nucleus and the spiral arms, and the ``low'' group is a field population. These populations apparently reflect the nuclear or arm and interarm ISM pressure regimes. For the nuclear region and the arm region, the substrate pressure is provided by the diffuse X-ray plasma. For the inter-arm region, the substrate pressure is not provided by the X-ray gas, and is more likely provided by turbulent motion in the molecular and warm phases of the ISM. }
\end{itemize}

\begin{acknowledgements}
This paper is based on observations with the NASA/ESA Hubble Space Telescope obtained at the Space Telescope Science Institute, which is operated by the Association of Universities for Research in Astronomy, Incorporated, under NASA contract NAS5-26555. It uses Early Release Science observations made by the WFC3 Scientific Oversight Committee.  We are grateful to the Director of the Space Telescope Science Institute for awarding  Director's Discretionary time for this program. Dopita acknowledges the support of the Australian Research Council (ARC) through Discovery  projects DP0984657 and DP0664434. This research has made use of the NASA/IPAC Extragalactic Database (NED) which is operated by  the Jet Propulsion Laboratory, California Institute of Technology, under contract with the National  Aeronautics and Space Administration.  This research has also made use of NASA's Astrophysics Data System, and of SAOImage DS9 \citep{joye03}, developed by the Smithsonian Astrophysical Observatory. We thank the anonymous referee for highly constructive criticism of the original version of our work, which greatly assisted us in the production of what has become a much better paper.
\end{acknowledgements}

\clearpage
\thispagestyle{empty}
\begin{deluxetable}{lcc} 
\tablecaption{Log of WFC3 Observations (Prop ID\# 11360).\label{tab:Obs}}
\tablehead{\colhead{Filter} & \colhead{Exposure (s)} &  \colhead{Date (2009)}}
\startdata
F336W & 600  & August 26 \\
F373N & $3 \times 800$ & August 20 \\
F438W & $3 \times 640$ & August 26 \\
F438W & 10 & August 26 \\
F487N & $3 \times 900$ & August 25 \\
F502N & $3 \times 828$ & August 26 \\
F555W & $3 \times 401$ & August 26 \\
F555W & 10 & August 20 \\
F657N & $4 \times 371$ & August 25 \\
F673N &  $2 \times 600$ & August 20 \\
F673N &  650 & August 20 \\
F814W &  $3 \times 401$ & August 26 \\
F814W & 10 & August 26 \\
\enddata
\end{deluxetable}
\clearpage

\clearpage
\thispagestyle{empty}
\begin{deluxetable}{lllccclll} 
\tabletypesize{\small}
\tablecaption{Catalogue of Outer ($R > 300$pc) SNR Candidates\label{tab:SNRlist1}}
\tablehead{\colhead{Name} & \colhead{RA (J2000)} & \colhead{Dec. (J2000)} & \colhead{D(arcsec)\tablenotemark{1}} &  \colhead{D(pc)} & \colhead{F(H$\alpha$) \tablenotemark{2}} & \colhead{F(H$\alpha)_{\rm B\&L}$\tablenotemark{3}} & \colhead{B\&L\#} & \colhead{Notes\tablenotemark{4}}}
\startdata
M83-SNR-1-01 & 13 36 58.712 & -29 51 00.52 & 1.14& 25.2& 1.1E-14 & & & X41, Strong [O III] \\
M83-SNR-1-02 & 13 36 58.885 & -29 52 26.02 & 0.92 & 20.4 &  &  &  & \\
M83-SNR-1-03 & 13 36 59.166 & -29 51 47.96 & 0.56 & 12.4 & 2.4E-15 &  & & X66 \\
M83-SNR-1-04 & 13 36 59.479 & -29 52 03.66 & 0.68 & 15.0 & 1.7E-14 & 1.64E-14 & BL31 & X45 \\
M83-SNR-1-05 & 13 36 59.725 & -29 50 32.88 & 1.18 & 26.1 & 1.7E-15 &  & \\ 
M83-SNR-1-06 & 13 37 00.068 & -29 52 08.43 & 0.52x0.86 & 15.3 & 9.0E-15 &  &  \\
M83-SNR-1-07 & 13 37 00.343 & -29 51 20.63 & 0.96x1.32 & 25.2 & 7.0E-15 & 4.20E-15 & BL33 & Strong [O III],[O II] \\
M83-SNR-1-08 & 13 37 00.672 & -29 52 21.75 & 0.96x1.44 & 26.5 &  &  &  & \\
M83-SNR-1-09 & 13 37 01.022 & -29 50 56.31 & 0.88 & 19.5 & 3.6E-15 &  &    \\
M83-SNR-1-10 & 13 37 01.076 & -29 51 41.63 &  0.56 & 12.4 & 4.9E-15 &  &  &  \\
M83-SNR-1-11 & 13 37 01.579 & -29 52 04.93 & 1.12 & 24.8 &  &  &  & \\
M83-SNR-1-12 & 13 37 01.729 & -29 51 13.47 & 1.40 & 31.0 & 2.1E-14 & 2.05E-14 & BL37 & X77,bipolar,strong [O III],[OII] \\
M83-SNR-1-13 & 13 37 02.002 & -29 52 50.03 & 2.2 & 48.7 & 2.1E-15 & 7.32E-15 & BL38  &  small H~II region in SW corner \\
M83-SNR-1-14 & 13 37 02.115 & -29 51 58.75 & 0.64x2.54 &  &  & 1.16E-14 & BL39 & Jet-like - very unusual morphology \\
M83-SNR-1-15 & 13 37 02.443 & -29 51 26.07 & 0.56 & 12.4 & 1.9E-14 & 2.28E-14 & BL41 &  X81, intense [O III], [O II] \\
M83-SNR-1-16 & 13 37 02.631 & -29 51 09.23 & 1.08 & 23.9 &  &  &  & small H~II region SW corner \\
M83-SNR-1-17 & 13 37 04.877 & -29 52 18.59 & 1.28 & 28.3 & 1.8E-15 &  &  & \\
M83-SNR-1-18 & 13 37 05.807 & -29 52 46.01 & 1.68 & 37.2 & 1.7E-15 &  &  & Very faint  \\
M83-SNR-1-19 & 13 37 06.435 & -29 50 25.19 & 1.58 & 34.9 & 1.4E-14 & 1.44E-14 & BL47 &  \\
M83-SNR-1-20 & 13 37 06.986 & -29 51 09.45 &  0.62 & 13.7 & 1.6E-15 &  &  &  \\
M83-SNR-1-21 & 13 37 07.686 & -29 51 09.93 & 0.38 & 8.4 &  &  &  \\
M83-SNR-1-22 & 13 37 07.471 & -29 51 33.23 & 1.32x1.58  & 32.1 & 2.3E-14 & 1.46E-14 & BL53 &  X105\\
M83-SNR-1-23 & 13 37 07.595 & -29 52 19.63 & 1.28 & 28.3 & 4.1E-15 & 5.80E-15 & BL55 &  \\
M83-SNR-1-24 & 13 37 07.852 & -29 52 41.93 & 1.26 & 27.9 & 1.4E-15 &  &  & R47? \\
M83-SNR-1-25 & 13 37 08.110 & -29 52 21.31 & 0.56x1.08 & 18.1 & 3.3E-15 &  &  & \\
M83-SNR-1-26 & 13 37 08.324 & -29 50 56.43 & 1.40 & 31.0 &  &  &  & \\
M83-SNR-1-27 & 13 37 08.490 & -29 52 02.35 & 2.30 & 50.9 &  &  &  & Bright H~II region to S, fainter to SW \\
M83-SNR-1-28 & 13 37 08.559 & -29 51 34.87 & 0.74 & 16.4 & 1.4E-14 & 1.19E-14 & BL58 & X109  \\
M83-SNR-1-29 & 13 37 08.667 & -29 52 42.69 & 1.26 & 27.9 & 6.1E-15 &  &  & \\
M83-SNR-1-30 & 13 37 08.656 & -29 51 53.57 & 0.72 & 15.9 & 5.9E-15 &  &  & \\ 
M83-SNR-1-31 & 13 37 08.696 & -29 51 33.29 & 1.84 & 40.7 & 1.3E-14 &  &  & \\
M83-SNR-1-32 & 13 37 08.753 & -29 51 37.41 & 1.26 & 27.9 & 8.6E-15 & 1.18E-14 & BL59 &  \\
M83-SNR-1-33 & 13 37 09.044 & -29 51 33.43 & 0.64 & 14.2 & 3.0E-15 &  &  & R52? \\
M83-SNR-1-34 & 13 37 09.227 & -29 51 34.06 & 0.60 & 13.3 & 2.5E-15 &  &  & R52? \\
M83-SNR-1-35 & 13 37 09.318 & -29 50 58.50 & 1.24 & 27.4 & 3.9E-14 &  &  & \\
M83-SNR-1-36 & 13 37 10.117 & -29 51 28.22 & 0.48 & 10.6 & 1.6E-14 &  &  & \\
M83-SNR-1-37 & 13 37 10.329 & -29 51 28.74 & 1.14 & 25.2 &  &  &  &  \\
M83-SNR-1-38 & 13 37 10.381 & -29 51 34.18 & 1.18 & 26.1 & 2.7E-15 &  & \\ 
M83-SNR-1-39 & 13 37 10.797 & -29 51 44.85 & 1.82 & 40.3 & 2.8E-15 & 5.08E-15 & BL60  & \\
M83-SNR-1-40 & 13 37 10.833 & -29 52 44.52 & 0.68 & 15.0 & 1.4E-15 &  &  \\
\enddata
\tablenotetext{1}{Typical measurement error $\pm0.08$ arcsec.}
\tablenotetext{2}{Units: erg cm$^{-2}$ s$^{-1}$, typical measurement error  $\pm15$\%.}
\tablenotetext{3}{From Table 5 of Blair \& Long (2004).}
\tablenotetext{4}{X: X-ray source number from Soria \& Wu (2003); R: radio source number from Maddox et al. (2006) using VLA data.}
\end{deluxetable}
\clearpage

\clearpage
\thispagestyle{empty}
\begin{deluxetable}{ccccccl} 
\tabletypesize{\small}
\tablecaption{Catalogue of Nuclear ($R<300$~pc) SNR Candidates\label{tab:SNRlist2}}
\tablehead{\colhead{Name} & \colhead{RA (J2000)} & \colhead{Dec. (J2000)} & \colhead{D(arcsec)\tablenotemark{1}} &  \colhead{D(pc)} 
& \colhead{F(H$\alpha$) \tablenotemark{2}} & \colhead{Notes\tablenotemark{3}}}
\startdata
M83 Nucleus & 13 37 00.871 & -29 51 55.97 & & &  \\
\hline
M83-SNR-N-01 & 13 37 00.039 & -29 52 02.08 & 0.20 & 4.4 & &  \\
M83-SNR-N-02 & 13 37 00.204 & -29 51 58.47 & 0.66x0.90 & 17.3 & &     \\
M83-SNR-N-03 & 13 37 00.214 & -29 52 06.27 & 0.68 & 15.0 & 4.9E-14 &  X53 \\
M83-SNR-N-04 & 13 37 00.335 & -29 52 05.52 & 0.32 & 7.1 & 2.0E-14  &  X54\\
M83-SNR-N-05 & 13 37 00.373 & -29 51 58.88 & 0.24x0.14 & 4.2 & &  Near SN1968L position.\\
M83-SNR-N-06 & 13 37 00.385 & -29 52 01.72 & 0.84 & 18.6 & &  \\
M83-SNR-N-07 & 13 37 00.404 & -29 52 06.33 & 0.52 & 11.5 & &   \\
M83-SNR-N-08 & 13 37 00.545 & -29 51 58.88 & 0.56 & 12.4 & 4.1E-13 & \\
M83-SNR-N-09 & 13 37 00.581 & -29 52 09.24 & 1.84 & 40.7 & &  \\
M83-SNR-N-10 & 13 37 00.609 & -29 51 59.65 & 0.76 & 16.8 & 3.9E-13 & X59? \\
M83-SNR-N-11 & 13 37 00.661 & -29 51 57.10 & 0.35x0.2 & 6.1 & &  \\
M83-SNR-N-12 & 13 37 00.701 & -29 51 59.89 & 0.86 & 19.0 & &    X59? \\
M83-SNR-N-13 & 13 37 00.757 & -29 52 06.49 & 0.44 & 9.7 & &  X62 \\
M83-SNR-N-14 & 13 37 00.929 & -29 51 54.25 & 0.88 & 19.5 & &  X63 \\
M83-SNR-N-15 & 13 37 00.932 & -29 51 55.93 & 0.18 & 4.0 & &   \\
M83-SNR-N-16 & 13 37 01.107 & -29 51 52.13 & 0.26x0.68 & 10.4  & &  X67\\
M83-SNR-N-17 & 13 37 01.172 & -29 51 57.19 & 0.52 & 11.5 & &   \\
M83-SNR-N-18 & 13 37 01.317 & -29 51 57.17 & 1.20 & 26.5 & &   \\
M83-SNR-N-19 & 13 37 01.593 & -29 52 02.29 & 0.96 & 21.2 & &   X74 \\
\enddata
\tablenotetext{1}{Typical measurement error $\pm0.08$ arcsec.}
\tablenotetext{2}{Units: erg cm$^{-2}$ s$^{-1}$, typical measurement error  $\pm15$\%.}
\tablenotetext{3}{X: source number from Soria \& Wu (2003); source crowding and coordinate uncertainties make some of these identifications tentative.}
\end{deluxetable}
\clearpage

\clearpage

\clearpage
\thispagestyle{empty}
\begin{deluxetable}{cccccl} 
\tabletypesize{\small}
\tablecaption{Extended [O~III] Nebulae in M83 WFC3 Field 1 \label{tab:strongOlist}}
\tablehead{\colhead{Name} & \colhead{RA (J2000)} & \colhead{Dec. (J2000)} & \colhead{D(arcsec)} &  \colhead{D(pc)} &  \colhead{Notes\tablenotemark{2}}}
\startdata
M83 Nucleus & 13 37 00.871 & -29 51 55.97 & &  \\
M83-SRC-1-O1 & 13 37 04.120 & -29 51 03.85 & 0.45x0.20 & 10x4 &   [O~III] arc in diffuse H$\alpha$.  \\
M83-SRC-1-O2 & 13 37 06.976 & -29 50 57.14 & 0.33 & 7.3 &   O-knots within faint H$\alpha$; possible O-rich SNR? \\
M83-SRC-1-O3 & 13 37 07.132 & -29 51 14.27 & 0.3x0.5 & 6.7x11.1 &   [O~III] knots within H$\alpha$ ring; O-rich SNR?; R45?\\
M83-SRC-1-O4 & 13 37 09.058 & -29 52 09.92 & 0.36 & 8.0 &   Young O-rich SNR in H~II region?\\
M83-SRC-1-O5 & 13 37 09.761 & -29 52 43.93 & 0.50 & 11.1 &   Strong [O~III] and H$\alpha$ ring; possible WR shell? \\
\enddata
\tablenotetext{2}{R: source number from Maddox et al. (2006) (VLA data).}
\end{deluxetable}
\clearpage

\clearpage
\thispagestyle{empty}
\begin{deluxetable}{ccccccccc} 
\tabletypesize{\small}
\tablecaption{Derived Parameters of M83 SNR.\label{tab:SNRparms}}
\tablehead{\colhead{Name} & \colhead{D} & \colhead{F(H$\alpha$) \tablenotemark{1}} & \colhead{log L \tablenotemark{2}} & \colhead{log $\tau_{\rm rad}$ \tablenotemark{3}} & \colhead{log S \tablenotemark{4}} &  \colhead{log $n_{\rm 0}$ \tablenotemark{5}} & \colhead{Class \tablenotemark{6}} & \colhead{$v_s({\rm max})$ \tablenotemark{7}}}
\startdata
M83-SNR-1-01 & 25.2 & 1.1E-14 & 39.31 & 4.19 & -0.95 & 1.04 & L & 460 \\
M83-SNR-1-02 & 20.4 &  &  &  &  &  &  & \\
M83-SNR-1-03 & 12.4 & 2.4E-15 & 38.65 & 4.85 & -0.99 &  0.99 & H & 640 \\
M83-SNR-1-04 & 15.0 & 1.7E-14 & 39.50 & 4.00 & -0.31 &  1.68 & H & 580 \\
M83-SNR-1-05 & 26.1 & 1.7E-15 & 38.50 & 5.00 & -1.79 &  0.20 & L & 450 \\
M83-SNR-1-06 & 15.3 & 9.0E-15 & 39.22 & 4.28 & -0.60 &  1.38 & H & 580 \\
M83-SNR-1-07 & 25.2 & 7.0E-15 & 39.11 & 4.39 & -1.14 &  0.84 & L & 460 \\
M83-SNR-1-08 & 26.5 & 2.2E-14 & 39.61 & 3.89 & -0.69 &  1.30 & L & 450 \\
M83-SNR-1-09 & 19.5 & 3.6E-15 & 38.83 & 4.68 & -1.21 &  0.78 & H & 520 \\
M83-SNR-1-10 & 12.4 & 4.9E-15 & 38.96 & 4.54 & -0.68 &  1.30 & H & 640 \\
M83-SNR-1-11 & 24.8 &  &  &  &  &  &  & \\
M83-SNR-1-12 & 31.0 & 2.1E-14 & 39.59 & 3.91 & -0.84 &  1.14 & L & 420 \\
M83-SNR-1-13 & 48.7 & 2.1E-15 & 38.59 & 4.91 & -2.24 & -0.25 & L & 340 \\
M83-SNR-1-14 &  &  &  &  &  &  & &   \\
M83-SNR-1-15 & 12.4 & 1.9E-14 & 39.55 & 3.95 & -0.09 &  1.89 & L & 640 \\
M83-SNR-1-16 & 23.9 & 1.7E-15 & 38.50 & 5.00 & -1.71 &  0.27 & L & 470 \\
M83-SNR-1-17 & 28.3 & 1.8E-15 & 38.52 & 4.98 & -1.83 &  0.15 & L & 440 \\
M83-SNR-1-18 & 37.2 & 1.7E-15 & 38.50 & 5.00 & -2.09 & -0.11 & L & 390 \\
M83-SNR-1-19 & 34.9 & 1.4E-14 & 39.42 & 4.09 & -1.12 &  0.86 & L & 400 \\
M83-SNR-1-20 & 13.7 & 1.6E-15 & 38.47 & 5.03 & -1.25 &  0.73 & H & 610 \\
M83-SNR-1-21 & 7.5   & 1.2E-15 & 38.33 & 5.17 &  -0.87&1.10 & H & 800 \\
M83-SNR-1-22 & 32.1 & 2.3E-14 & 39.63 & 3.87 & -0.84 &  1.15 & L & 410 \\
M83-SNR-1-23 & 28.3 & 4.1E-15 & 38.88 & 4.62 & -1.47 &  0.51 & L & 440 \\
M83-SNR-1-24 & 27.9 & 1.4E-15 & 38.42 & 5.09 & -1.93 &  0.05 & L & 440 \\
M83-SNR-1-25 & 18.1 & 3.3E-15 & 38.79 & 4.71 & -1.18 &  0.80 & H & 540 \\
M83-SNR-1-26 & 31.0 &  &  &  &  &  &  & \\
M83-SNR-1-27 & 50.9 &  &  &  &  &  &  & \\
M83-SNR-1-28 & 16.4 & 1.4E-14 & 39.42 & 4.09 & -0.47 &  1.52 & H & 560 \\
M83-SNR-1-29 & 27.9 & 6.1E-15 & 39.05 & 4.45 & -1.29 &  0.69 & L & 440 \\
M83-SNR-1-30 & 15.9 & 5.9E-15 & 39.04 & 4.46 & -0.82 &  1.17 & H & 570 \\
M83-SNR-1-31 & 40.7 & 1.3E-14 & 39.38 & 4.12 & -1.29 &  0.69 & H & 470 \\
M83-SNR-1-32 & 27.9 & 8.6E-15 & 39.20 & 4.30 & -1.14 &  0.84 & L & 440 \\
M83-SNR-1-33 & 14.2 & 3.0E-15 & 38.75 & 4.75 & -1.01 &  0.97 & L & 600 \\
M83-SNR-1-34 & 13.3 & 2.5E-15 & 38.67 & 4.83 & -1.03 &  0.95 & H & 620 \\
M83-SNR-1-35 & 27.4 & 3.9E-14 & 39.86 & 3.64 & -0.47 &  1.52 & L & 440 \\
M83-SNR-1-36 & 10.6 & 1.6E-14 & 39.47 & 4.03 & -0.03 &  1.95 & H & 680 \\
M83-SNR-1-37 & 25.2 & 1.6E-14 & 39.47 & 4.03 & -0.78 &  1.20 & L & 460  \\
M83-SNR-1-38 & 26.1 & 2.7E-15 & 38.70 & 4.80 & -1.59 &  0.40 & L & 450 \\
M83-SNR-1-39 & 40.3 & 2.8E-15 & 38.72 & 4.78 & -1.95 &  0.04 & L & 370 \\
M83-SNR-1-40 & 15.0 & 1.4E-15 & 38.42 & 5.09 & -1.39 &  0.59 & L & 580 \\
M83-SNR-1-N02 & 17.3 & 1.0E-14 & 39.27 & 4.23 & -0.66 &  1.32 & H & 550 \\
M83-SNR-1-N03 & 15.0 & 4.9E-14 & 39.96 & 3.54 &  0.15 &  2.14 & H & 580 \\
M83-SNR-1-N04 & 7.1    & 2.0E-14 & 39.56 & 3.94 &  0.41 &  2.39 & H & 820 \\
M83-SNR-1-N07 & 11.5 &  &  &  &  &  &  & \\
M83-SNR-1-N08 & 12.4 & 4.1E-13 & 40.88 & 2.62 &  1.24  &  3.23 & H & 640 \\
M83-SNR-1-N10 & 16.8 & 3.9E-13 & 40.85 & 2.65 &  0.95  &  2.93 & H & 560 \\
M83-SNR-1-N16 & 10.4 &  &  &  &  &  &  & \\
M83-SNR-1-N18 & 26.5 &  &  &  &  &  &  & \\
Meas. Error \tablenotemark{8} & $\pm 1.8$ & $\pm 10$\%  & $\pm 0.15$ & $\pm 0.15$ & $\pm 0.20$ & $\pm 0.20$ &  & $\pm 25$\% \\
\enddata
\tablenotetext{1}{erg~cm$^{-2}$~s$^{-1}$.}
\tablenotetext{2}{Total luminosity: erg~s$^{-1}$.}
\tablenotetext{3}{Radiative age: yr.}
\tablenotetext{4}{Mean surface energy flux of radiative shocks: erg~cm$^{-2}$~s$^{-1}$.}
\tablenotetext{5}{Inferred pre-shock density: cm$^{-3}$.}
\tablenotetext{6}{Density class: high--H, low--L.}
\tablenotetext{7}{Maximum shock velocity: km~s$^{-1}$.}
\tablenotetext{8}{Errors given here are typical, and based upon measurement errors only. Errors in the assumptions used may produce much larger systematic errors. These can be estimated from the formulae given in section \ref{Parms}.}
\end{deluxetable}
\clearpage

\end{document}